\documentclass{article}
\usepackage{graphicx}
\newcommand{\myscalebox}[1]{\scalebox{0.43}[0.43]{#1}}
\newenvironment{myalgorithm}[1]
{
\begin{tabbing} xx \= xx \= xx \= xx \= xx \= xx \= xx \= xx \kill
 {\bf algorithm #1}\\
 {\bf begin}\\
}
{
 {\bf end}\\
 \end{tabbing}
}

\begin{document}
\title{Minimal vertex covers on finite-connectivity random graphs 
   -- a hard-sphere lattice-gas picture}
\author{Martin Weigt and Alexander K. Hartmann \\
Institute for Theoretical Physics,\\
University of G\"ottingen, Bunsenstr. 9, 37073 G\"ottingen, Germany\\
E-mail: {\tt hartmann/weigt@theorie.physik.uni-goettingen.de}
}

\date{\today}
\maketitle

\begin{abstract}
The minimal vertex-cover (or maximal independent-set) problem is 
studied on random graphs of
finite connectivity. Analytical results are obtained by a mapping to
a lattice gas of hard spheres of (chemical) radius one, and they are
found to be in excellent agreement with numerical simulations. 
We give a detailed description of the replica-symmetric phase,
including the size and the entropy of the minimal vertex covers,
and the structure of the unfrozen component which is found to
percolate at connectivity $c\simeq 1.43$. The replica-symmetric 
solution breaks down at $c=e\simeq 2.72$. We give a simple
one-step replica symmetry broken solution, and discuss the problems
in interpretation and  generalization of this solution.
\end{abstract}


\section{Introduction}\label{sec:intro}

The last few years have seen an increasing interest of theoretical
computer scientists, mathematicians and, more recently, of statistical
physicists in random combinatorial optimization and decision problems,
see {\it e.g.} \cite{AI,TCS}.  Traditional complexity theory
\cite{GaJo} characterizes combinatorial problems with respect to the
worst-case dependence of solution times for algorithms on the problem
size, or, more precisely, on the memory size needed to encode a
problem. Some of the most challenging problems are collected in the
class of {\it NP-complete problems}: In such problems a potential
solution can be verified (or falsified) very effectively in polynomial
time, whereas the search for a solution among the exponential number
of candidates becomes very slow due to entropic reasons. The
completeness property refers to the fact that once an effective, {\it
i.e.} polynomial algorithm is found for any NP-complete problem, it
can be modified to effectively solve every other such problem.  The
question whether or not such algorithms can be constructed is however
still open, and belongs to the important open questions of modern
mathematics. Famous members of the class of NP-complete problems are
{\it e.g.}  Boolean satisfiability, number partitioning, vertex cover,
or the traveling-salesman problem.

This worst-case classification does however give no information on
typical solution times. For almost ten years now, randomized
optimization and decision problems have been studied therefore, for an
overview see the special issues \cite{AI,TCS}. It was realized, that
exponentially longest solution times typically appear when the
problems are situated at phase boundaries and therefore critically
constraint \cite{MiSeLe}.

Due to the analogy between such combinatorial optimization problems 
and statistical-mechanics models with discrete degrees of freedom at 
low temperature, many methods developed in physics can be applied
directly to theoretical computer science. This was done {\it e.g.}
for Boolean satisfiability \cite{MoZe,nature,BiMoWe,RiWeZe}, for
number partitioning \cite{Me}, the traveling-salesman problem
\cite{MePa}, for Euclidean matching \cite{HoBoMa} and recently also 
for vertex cover \cite{WeHa}. Also the relations between phase
transitions and the appearance of hardest instances was recently
analyzed for specific algorithms using statistical-mechanics methods 
\cite{CoMo,WeHa2}. 

In this paper, we give a detailed description of the statistical
mechanics approach to minimal vertex covers on finite connectivity
random graphs. For this reason, the model will be mapped to a random
lattice gas of hard spheres of radius one.

The plan of the paper is the following. In the next section we define
the model and give an overview over some rigorously known results. In
section \ref{sec:latticegas} the model is shown to be equivalent to a
hard-sphere lattice gas. Section \ref{sec:num} explains the numerical
methods used to check the analytical results. The latter are based on
the replica approach presented in sections
\ref{sec:replica}-\ref{sec:1rsb}, starting with a general calculation
of the replicated free energy.  In section \ref{sec:rs}, the most
important results are presented: the size, entropy and structure of
minimal vertex covers are described in a replica-symmetric approach,
whereas the simplest one-step replica symmetry broken solution is
explained in section \ref{sec:1rsb}. The paper is closed by a
concluding section. Several technical details are delegated to three
appendices.


\section{The model}\label{sec:model}

In this section, we will introduce the terminology and some rigorously 
known results about vertex cover and related problems.

\subsection{Vertex cover and related problems}\label{sec:vc}

Let us start with the definition of vertex covers. We consider a graph
$G=(V,E)$ with $N$ vertices $i\in \{1,2,...,N\}$ and undirected edges 
$\{i,j\}\in E\subset V\times V$ connecting pairs of vertices. Please
note that $\{i,j\}$ and $\{j,i\}$ both denote the same edge.

Definition 1: {\it A vertex cover $V_{vc}$ is a subset $V_{vc}\subset
  V$ of vertices such that for all edges $\{i,j\}\in E$ at least one of
the endpoints is in $V_{vc}$, i.e. $i\in V_{vc}$ or $j\in V_{vc}$.}

Later on also subsets $V_{vc}$ are considered, which are not
covers. Anyway, we call all vertices in $V_{vc}$ {\em covered}, all
others {\em uncovered}. Also edges from $E\cap (V_{vc}\times V \cup V
\times V_{vc})$ are called covered.  This means that if $V_{vc}$ is a
vertex cover, all edges are covered.

We will study the minimal vertex-cover problem, which consists in
finding a vertex cover $V_{vc}$ of minimal cardinality, and calculate 
the minimal fraction $x_c(G)=|V_{vc}|/N$ needed to cover the whole 
graph.

This problem is equivalent to other optimization problems:
\begin{itemize}
\item An {\it independent set} is a subset of vertices which are
  pairwise disconnected in the graph $G$. Due to the above-mentioned
  properties, any set $V\setminus V_{vc}$ thus forms an independent
  set, and maximal independent sets are complementary to minimal
  vertex covers.
\item A {\it clique} is a fully connected subset of vertices, and thus
  an independent set in the complementary graph $\overline{G}$ where
  vertices $i$ and $j$ are connected whenever $\{i,j\}\notin E$ and
  vice versa.
\end{itemize}

\subsection{Random graphs}\label{sec:rg}

In order to speak of median or average cases, and of phase
transitions, we have to introduce a probability distribution over
graphs. This can be done best by using the concept of {\it random
graphs} as already introduced about 40 years ago by Erd\"os and
R\'enyi \cite{ErRe}.  A random graph $G_{N,p}$ is a graph with $N$
vertices $V=\{1,...,N\}$, any pair of vertices is connected randomly
and independently by an edge with probability $p$. So the expected
number of edges becomes $p {N\choose 2} = pN^2/2+O(N)$, and the
average connectivity of a vertex equals $p(N-1)$.

The regime we are interested in are however {\it finite-connectivity 
graphs} having $p=c/N$ with constant $c$ in the large-$N$ limit. Then 
the average connectivity $c+O(N^{-1})$ stays finite. In this case, we
also expect the size of minimal vertex covers to depend only on $c$,
$x_c(G)=x_c(c)$ for almost all random graphs $G_{N,c/N}$.

Here we want to review shortly some of the fundamental results on
random graphs which were already described in \cite{ErRe}, and which 
are important for the following sections:

The first point we want to mention is the distribution of
connectivities (or vertex degrees) $d$, in the limit $N\to\infty$ it
is given by a Poisson-distribution with mean $c$:
\begin{equation}
  \label{eq:Poisson}
  Po_{c}(d) = e^{-c} \frac{(c)^d}{d!}\ . 
\end{equation}
A second point which is important for the understanding of the following 
is the component structure. For $c<1$, {\it i.e.} if the vertices have 
in average less than one neighbor, the graph $G_{N,c/N}$ is built up from 
connected components containing up to $O(\ln N)$ vertices. The 
probability that a component is a specific tree $T_k$ of $k$ vertices 
is given by
\begin{equation}
  \label{components}
  \rho (k) = e^{-ck} \frac{(c)^{k-1}}{k!}\ ,
\end{equation}
and is equal for all $k^{k-2}$ distinct trees. As the fraction of 
vertices which are collected in finite trees is 
$\sum_{k=1}^{\infty} \rho (k) k^{k-2} k = 1$ for all $c<1$, in this
case almost all vertices are collected in such trees. For $c>1$ a
giant component appears which contains a finite fraction of all 
vertices. $c=1$ is therefore called the {\em percolation threshold}.

For a complete introduction to random graphs see the book by
Bollobas \cite{Bo}.

\subsection{Rigorously known bounds}\label{sec:bound}

In this subsection we are going to present some previously known 
rigorous bounds on $x_c(c)$. A general one for arbitrary, 
{\it i.e.} non-random graphs $G$ was given by Harant \cite{Ha}
who generalized an old result of Caro and Wei \cite{CaWe}. Translated 
into our notation, he showed that
\begin{equation}
  \label{bound_harant}
 x_c(G)\leq 1-\frac{1}{N}\frac{\left(\sum_{i\in V}\frac{1}{d_i+1} 
                                   \right)^2}{
                  \sum_{i\in V}\frac{1}{d_i+1} - \sum_{(i,j)\in E} 
                      \frac{(d_i-d_j)^2}{(d_i+1)(d_j+1)}}
\end{equation}
where $d_i$ is the connectivity (or degree) of vertex $i$. Using the 
distribution
(\ref{eq:Poisson}) of connectivities and its generalization to pairs of 
connected vertices, this can easily be converted into an upper bound
on $x_c(c)$ which holds almost surely for $N\to\infty$.

The vertex cover problem or the above-mentioned related problems
were also studied in the case of random graphs, and even completely 
solved in the case of infinite connectivity graphs, where any edge
is drawn with finite probability $p$, such that the expected number of 
edges is $p {N\choose 2}=0(N^2)$. There the minimal VC has cardinality
$(N-2\ln_{1/(1-p)}N-O(\ln \ln N))$ \cite{BoEr}. Bounds in the 
finite-connectivity region of random graphs with $N$ vertices and $cN$
edges were given by Gazmuri \cite{Ga}. He showed that
\begin{equation}
  \label{bound_gazmuri}
  x_l(c) < x_c(c) < 1- \frac{\ln c}{c}
\end{equation}
where the lower bound is given by the unique solution of
\begin{equation}
  \label{low}
  0=  x_l(c) \ln  x_l(c) + (1- x_l(c)) \ln (1- x_l(c))
      - \frac{c}{2} (1- x_l(c))^2\ .
\end{equation}
This bound coincides with the so-called
annealed bound in statistical physics. The correct asymptotics for
large $c$ was given by Frieze \cite{Fr}:
\begin{equation}
  \label{asympt}
  x_c(c) = 1 - \frac{2}{c}(\ln c - \ln\ln c +1 - \ln2 )
   +o(\frac{1}{c})
\end{equation}
with corrections of $o(1/c)$ decaying faster than $1/c$.


\section{Equivalence to a hard-sphere lattice gas}\label{sec:latticegas}

After having introduced the problem in mathematical terms, we are now
going to connect it to a statistical-mechanics model, more precisely
to a lattice gas of hard spheres of chemical radius one.

Any subset $U\subset V$ of the vertex set can be encoded by a
configuration of $N$ binary variables:
\begin{equation}
  \label{eq:mapping}
  x_i := \left\{
      \begin{array}{lll}
       0 & \mbox{if} & i\in U\\
       1 & \mbox{if} & i\notin U
      \end{array}
       \right.
\end{equation}
The strange choice of setting $x_i$ to zero for vertices in $U$
becomes clear if we look at the vertex cover constraint: An edge is
covered by the elements in $U$ iff at most one of the two end-points
has $x=1$. So the variables $x_i$ can be interpreted as occupation
numbers of vertices by the center of a particle. The covering
constraint translates into a hard sphere constraint: If a vertex is
occupied, {\it i.e.} $x_i=1$, then all neighboring vertices have to be
empty. We thus introduce an indicator function
\begin{equation}
  \label{eq:chi}
  \chi(x_1,...,x_N)=\prod_{\{i,j\}\in E} (1-x_i x_j)
\end{equation}
which is one whenever $\vec x = (x_1,...,x_N)$ corresponds to a vertex
cover, and zero else. Having in mind this interpretation, we may write
down the grand partition function
\begin{equation}
  \label{eq:xi}
  \Xi = \sum_{\{x_i=0,1\}} \exp\left(\mu \sum_i x_i\right) \ \chi(\vec x) 
\end{equation}
with $\mu$ being a chemical potential which can be used to control the
particle number, or the cardinality of $U$. 

For regular lattices, this model is well studied as a lattice model
for the fluid-solid transition, for an overview and the famous
corner-transfer matrix solution of the two-dimensional hard-hexagon 
model by R. Baxter see his book \cite{Ba}. Recently, lattice-gas
models with various kinds of disorder have been considered in
connection to glasses \cite{glass1,glass2,glass3} and granular matter 
\cite{granular1,granular2,granular3,granular4,granular5,granular6}. 

Denoting the grand canonical average as
\begin{equation}
  \label{eq:mean}
  \langle f(\vec x) \rangle_\mu=\Xi^{-1} \sum_{\{x_i=0,1\}} 
  \exp\left(\mu \sum_i x_i\right) \ \chi(\vec x)\ f(\vec x)
\end{equation}
we can calculate the average occupation density
\begin{equation}
  \label{eq:N}
  \nu(\mu) = \frac{1}{N}\langle \sum_i x_i \rangle_\mu
  = \frac{\partial}{\partial\mu} \frac{\ln\Xi}{N}\ ,
\end{equation}
and the corresponding entropy density is given by a Legendre 
transform of $\ln \Xi$,
\begin{equation}
  \label{eq:entropy}
  s(\nu(\mu)) = \left( 1- \mu\frac{\partial}{\partial\mu}\right) 
           \frac{\ln\Xi}{N} ,
\end{equation}
where the thermodynamic limit $N\to\infty$ is implicitly assumed.
The entropy of vertex covers of cardinality $xN$ thus reads
\begin{equation}
  \label{eq:vcentr}
  s_{vc}(x) = s(1-x)
\end{equation}

Minimal vertex covers correspond to densest particle packings. 
Considering the weights in (\ref{eq:xi}), it becomes obvious that the
density $\rho(\mu)$ is an increasing function of the chemical
potential. Densest packings, or minimal vertex covers, are thus
obtained in the limit $\mu\to\infty$:
\begin{equation}
  \label{eq:mulimit}
  x_c(c) = 1-\lim_{\mu\to\infty} \nu(\mu)\ .
\end{equation}


\section{Numerical methods}\label{sec:num}

Before explicitly following this strategy in the special case of
random graphs, we are going to present our numerical methods. So
we can later-on  directly compare all analytical results to numerical
data.

All numerical results were obtained by exact enumerations. For large
average connectivities $c\ge 4$ a branch-and-bound algorithm was
applied, while for small average connectivities a divide-and-conquer
technique is more appropriate. Since some readers may not be familiar
with combinatorial optimization algorithms, the methods are explained
in detail.  Before presenting the two procedures, we first introduce a
fast heuristic, which is used within both methods. The heuristic can
applied stand-alone as well. In this case only an approximation of the
true minimum vertex cover is calculated, which is found to differ only
by a few percent from the exact value. All methods have been
implemented via the help of the LEDA library \cite{leda99} which
offers many useful data types and algorithms for linear algebra and
graph problems.

The basic idea of the heuristic is to cover as many edges as possible
by using as few vertices as necessary. Thus, it seems favorable to
cover vertices with a high degree. This step can be iterated, while
the degree of the vertices is adjusted dynamically by removing edges
and vertices which are covered.  This leads to the following
algorithm, which returns an approximation of the minimum vertex cover
$V_{vc}$, the size $|V_{vc}|$ is an upper bound of the true minimum
vertex-cover size:
\begin{myalgorithm}{min-cover($G$)}
\> initialize $V_{vc}=\emptyset$;\\
\> {\bf while} there are uncovered edges {\bf do}\\
\> {\bf begin}\\
\>\> take one vertex $i$ with the largest current degree $d_i$;\\
\>\> mark $i$ as covered: $V_{vc} = V_{vc} \cup \{i\}$;\\
\>\> remove all incident edges $\{i,j\}$ from $E$;\\
\>\> remove vertex $i$ from $V$;\\
\> {\bf end};\\
\> return($V_{vc}$);\\
\end{myalgorithm}

In Fig. \ref{figHeuristicBad} a simple example is presented, where the
heuristic fails to find the true minimal vertex cover.  First the
algorithm covers the root vertex of degree 3. Thus, additionally 3
vertices have to be subsequently covered, i.e. the heuristic covers 4
vertices.  But, the minimum vertex cover has only size 3, as indicated
in Fig. \ref{figHeuristicBad}.

So far we have presented a simple heuristic to  find approximations of
minimum vertex covers, which will be
part of the exact algorithms, which we have been applied to obtain all
numerical results presented in this work. Next, two exact algorithms
are explained: divide-and-conquer and branch-and-bound.

The basic idea of both methods is as follows: as each vertex is either
covered or uncovered, there are $2^N$ possible configurations which
can be arranged as leafs of a binary (backtracking) tree. At each
node, the two subtrees represent the subproblems where the
corresponding vertex is either {\em covered} (``left subtree'') or
{\em uncovered} (``right subtree'').  Vertices, which have not been
touched at a certain level of the tree are said to be {\em free}.
Both algorithms do not descent further into the tree, when a cover has
been found, i.e. when all edges are covered. Then the search continues
in higher levels of the tree (backtracking) for a cover which has
possibly a smaller size. Since the number of nodes in a tree grows
exponentially with system size, algorithms which are based on
backtracking trees have a running time which may grow exponentially
with the system size. This is not surprising, since the minimal-VC 
problem is NP-hard, so all exact methods exhibit an exponential growing
worst-case time complexity.
 
To decrease running time, both algorithms make use of the fact, that
only full vertex covers are to be obtained. Therefore, when a vertex
$i$ is marked {\em uncovered}, all neighboring vertices can be {\em
covered} immediately. Concerning these vertices, only the left
subtrees are present in the search tree.

The divide-and-conquer \cite{aho74} approach is based on the fact that
a minimum VC of a graph, which consists of several independent
connected components, can be obtained by combining the minimum covers
of the components. Thus, the full task can be split into several
independent tasks. This strategy can be repeated at all levels of the
backtracking tree. At each level, the edges which have been covered
can be removed from the graph, so the graph may split into further
components. As a consequence, below the percolation threshold, where
the size of the largest components is of the order $O(\ln N)$, the
algorithm exhibits a polynomial running time. Summarizing, the
divide-and-conquer approach reads as follows, the given subroutine is
called for each component of the graph separately, it returns the size
of the minimum vertex cover.  Initially all vertices have state {\em
free}:

\begin{myalgorithm}{divide\_and\_conquer(G)}
\> take one {\em free} vertex $i$ with the largest current degree $d_i$;\\
\>\\
\> mark $i$ as {\em covered}; \>\>\>\>\>\>{\bf comment} left subtree\\
\> $size_1:=1$;\\
\> remove all incident edges $\{i,j\}$ from $E$;\\
\> calculate all connected components $\{C_i\}$ of graph built by {\it
  free} vertices;\\
\> {\bf for} all components $C_i$ {\bf do}\\
\>\>  $size_1 := size_1 + $ {\bf divide\_and\_conquer}($C_i$);\\
\> insert all edges $\{i,j\}$ which have been removed;\\
\>\\
\> mark $i$ as {\em uncovered}; {\bf comment} right subtree;\\
\> $size_2:=0$;\\
\> {\bf for} all neighbors $j$ of $i$ {\bf do}\\
\> {\bf begin}\\
\>\> mark $j$ as {\em covered}\\
\>\>  remove all incident edges $\{j,k\}$ from $E$;\\
\> {\bf end}\\
\> calculate all connected components $\{C_i\}$;\\
\> {\bf for} all components $C_i$ {\bf do}\\
\> {\bf } $size_2 := size_2 + $ {\bf divide\_and\_conquer}($C_i$);\\
\> {\bf for} all neighbors $j$ of $i$ {\bf do}\\
\>\> mark $j$ as {\em free}\\
\> insert all edges $\{j,k\}$ which have been removed;\\
\>\\
\> mark $i$ as {\em free};\\
\> {\bf if} $size_1 < size_2$ {\bf then}\\
\>\> {\bf return}($size_1$);\\
\> {\bf else}\\
\>\> {\bf return}($size_2$);\\
\end{myalgorithm}

The algorithm can be easily extended to record the cover sets as well
or to calculate the degeneracy. In Fig. \ref{figExampleDivide} an
example of the operation is given.
The algorithm is able to treat large graphs deep in the percolating
regime. We have calculated for example minimum vertex covers for
graphs of size $N=560$ with average connectivity $c=1.3$

For average connectivities larger than $4$, the divide-and-conquer
algorithm is too slow, because the graph only rarely splits into
several components. Then a branch-and-bound approach
\cite{Lawler66,Tarjan77,Shindo90} is favorable. It differs from the
previous method by the fact that no independent components of the
graph are calculated. Instead, some subtrees of the backtracking tree
are omitted by introducing a {\em bound:} This is achieved by storing
always the size $best$ of the smallest vertex cover found so far
(initially $N$) and recording the number $X$ of vertices which have
bee {\em covered} in higher levels of the tree. Additionally, always a
table of {\em free} vertices ordered in descending current degree
$d_i$ is kept. Thus, to achieve a better solution, at most $F=best-X$
vertices can be {\em covered}. This means, it is not possible to cover
more edges, than given by the sum $D=\sum_{l=1}^F d_l$ of the $F$
highest degrees in the table of vertices, {\it i.e.} if some edges
remain uncovered, the corresponding subtree can be omitted for
sure. Please note that in the case that some edges are running between
the $F$ vertices of highest current degree, then a subtree may be
entered, even if it contained no smaller cover.

The algorithm can be summarized as follows below. The size of the
smallest covered is stored in $best$, which is passed by reference
(i.e. the variable, not its value is passed). The number of covered
vertices is stored in variable $X$, please remember $G=(V,E)$:

\begin{myalgorithm}{branch\_and\_bound($G$, $best$, $X$)}
\> {\bf if} all edges are {\em covered} {\bf then}\\
\> {\bf begin}\\
\>\> {\bf if} $X<best$ {\bf then}\\
\>\>\> $best:=X$;\\
\>\> {\bf return};\\
\> {\bf end};\\
\> calculate $F=best-X$; $D=\sum_{l_1}^F d_l$;\\
\> {\bf if} $D <$ number of uncovered edges {\bf then}\\
\>\> {\bf return}; \>\>\>\>\> {\bf comment} bound; \\
\> take one {\em free} vertex $i$ with the largest current degree $d_i$;\\
\>\\
\> mark $i$ as {\em covered}; \>\>\>\>\>\> {\bf comment} left subtree\\
\> $X:=X+1$;\\
\> remove all incident edges $\{i,j\}$ from $E$;\\
\> {\bf branch\_and\_bound}($G$, $best$, $X$);\\
\> insert all edges $\{i,j\}$ which have been removed;\\
\> $X:=X-1;$\\
\>\\
\> {\bf if} ($X>$ number of current neighbors) {\bf then}\\
\> {\bf begin} \>\>\>\>\>\> {\bf comment} right subtree;\\
\>\> mark $i$ as {\em uncovered};\\ 
\>\> {\bf for} all neighbors $j$ of $i$ {\bf do}\\
\>\> {\bf begin}\\
\>\>\> mark $j$ as {\em covered};\\
\>\>\> $X:=X+1$;\\
\>\>\>  remove all incident edges $\{j,k\}$ from $E$;\\
\>\> {\bf end};\\
\>\> {\bf branch\_and\_bound}($G$, $best$, $X$);\\
\>\> {\bf for} all neighbors $j$ of $i$ {\bf do}\\
\>\> {\bf begin}\\
\>\>\> mark $j$ as {\em free};\\
\>\>\> $X:=X-1$;\\
\>\> {\bf end};\\
\>\> insert all edges $\{j,k\}$ which have been removed;\\
\> {\bf end};\\
\>\\
\> mark $i$ as {\em free};\\
\> {\bf return};\\
\end{myalgorithm}

For every calculation of the bound one has to access the $F$ vertices
of largest current connectivity. Thus it it favorable to implement the
table of vertices as two arrays $v_1,v_2$ of sets of vertices.  The
arrays are indexed of the current degree of the vertices. The sets of
the first array $v_1$ contains the $F$ {\em free} vertices of largest
current degree, while the other array contains all other {\em free}
vertices. Every time a vertex changes its degree, it is moved to
another set, and eventually even changes the array. Also, in case the
mark of a vertex changes it may be entered in or removed from an
array, possibly the smallest degree vertex of $v_1$ is moved to $v_2$
or vice versa. Since we are treating graphs of finite average
connectivity, this implementation ensures that the running time spent
in each node of the graph is almost constant. For the sake of clarity,
we have omitted the update operation for both arrays from the
algorithmic presentation.

Although our algorithm is very simple, in the regime $4<c<10$ random
graphs up to size $N=140$ could be treated. It is difficult to compare
the branch-and-bound algorithm to state-of-the-art algorithms
\cite{Shindo90,lueling1992} because they are usually tested on a
different graph ensemble where each edge appears with a certain
probability, independently of the graph size (high-connectivity
regime).  Nevertheless, in the literature usually graphs with up to
200 vertices are treated, which is slightly larger than the systems
considered here.  But our algorithm has the advantage that it is easy
to implement. Furthermore, it can be easily modified to study more
general questions, see \cite{WeHa}


\section{Replica approach}\label{sec:replica}

After having introduced our numerical methods, we go back to the
statistical-mechanics approach displayed in section
\ref{sec:latticegas}.  The main problem in handling the grand
partition function (\ref{eq:xi}) is caused by the disorder due to the
random structure of the underlying graph, {\it i.e.} of the edge set
$E$. To calculate typical properties we therefore have to evaluate the
disorder average of $\ln\Xi$ over the random graph ensemble. This can
be achieved by the replica trick \cite{MePaVi},
\begin{equation}
  \label{eq:rep}
  \overline{\ln \Xi} = \lim_{n\to 0} \frac{\overline{\Xi^n}-1}{n}
\end{equation}
where the over-bar denotes the disorder average over the random-graph
ensemble. Taking $n$ to be a positive integer at the beginning, we may 
replace the original system by $n$ identical copies (including
identical edge sets). In this case, the disorder average is easily
obtained, and the $n\to 0$ limit has to be achieved later by
analytically continuing in $n$. We may thus write, with $n$ being a
natural number,
\begin{equation}
  \label{eq:xin}
  \overline{\Xi^n}=\sum_{\{x_i^a\}} \exp\left(\mu\sum_{i,a}x_i^a 
           \right) \overline{\prod_{\{i,j\}\in E}\prod_{a=1}^{n}
          (1-x_i^ax_j^a)}
\end{equation}
with $a$ denoting the replica index which runs from $1$ to $n$.
Putting edges independently with probability $c/N$ results in
\begin{eqnarray}
  \label{eq:average}
  \overline{\Xi^n}&=&\sum_{\{x_i^a\}}\exp\left(\mu\sum_{i,a}x_i^a 
           \right)\prod_{i<j}\left[ 1
    -\frac{c}{N}+ \frac{c}{N}\prod_a(1-x_i^a x_j^a) \right]\nonumber\\
 &=& \sum_{\{x_i^a\}}\exp\left(\mu\sum_{i,a}x_i^a 
      -\frac{cN}{2}+\frac{c}{2N} \sum_{i,j} \prod_a(1-x_i^a x_j^a) 
      + O(1) \right) .
\end{eqnarray}
Following the ideas of \cite{Mo}, we introduce $2^n$ order parameters
\begin{equation}
  \label{eq:op}
  c(\vec\xi) = \frac{1}{N} \sum_i \prod_a \delta_{\xi^a,x_i^a}
\end{equation}
as the fraction of vertices showing the replicated variable
$\vec\xi\in \{0,1\}^n$. The exponent in the last line of 
(\ref{eq:average}) obviously depends only on this quantity. Using
Stirling´s formula for the number $N!/\prod_{\vec\xi}(c(\vec\xi)N)!$
of configurations of the $\{x_i^a\}$ having the same $c(\vec\xi)$, 
we find
\begin{equation}
  \label{eq:freeener}
  \overline{\Xi^n}=\int {\cal D}c(\vec\xi) \exp\left\{
      N\left(-\sum_{\vec\xi} c(\vec\xi)\ln c(\vec\xi) -\frac{c}{2}
        +\mu \sum_{\vec\xi,a}c(\vec\xi)\xi^a +\frac{c}{2}
        \sum_{\vec\xi,\vec\zeta}c(\vec\xi)c(\vec\zeta)
        \prod_a(1-\xi^a\zeta^a) \right) \right\}
\end{equation}
where the integration is over all normalized distributions
$c(\vec\xi)$, {\it i.e.} $\sum_{\vec\xi} c(\vec\xi)=1$. In the
large-$N$ limit, the integration can be solved by the saddle-point
method. The saddle-point equation can be obtained by variation of
the exponent in (\ref{eq:freeener}) with respect to all allowed
$c(\vec\xi)$:
\begin{equation}
  \label{eq:sp}
  c(\vec\xi) = \exp\left\{ -\lambda +\mu\sum_a\xi^a+c\sum_{\vec\zeta}
    c(\vec\zeta) \prod_a(1-\xi^a\zeta^a)  \right\}\ .
\end{equation}
$\lambda$ is a Lagrange multiplier introduced in order to guarantee
the normalization of $c(\vec\xi)$. For $n\to 0$, it will tend to the
connectivity $c$. Before we can however calculate this limit, we have
to introduce some ansatz for $c(\vec\xi)$ as even the dimensionality
of $c(\xi)$ still depends on $n$. In the next section, we are going to
use the simplest-possible, {\it i.e.} the replica-symmetric ansatz. As
this ansatz is found to be valid only for a finite connectivity range,
we also include one step of replica-symmetry breaking in the over-next
section. 

\section{The replica-symmetric solution}\label{sec:rs}

\subsection{The replica-symmetric ansatz}

As already explained, we are now using the so-called replica-symmetric
ansatz, which in our case assumes that the order parameter
$c(\vec\xi)$ depends on $\vec\xi$ only via $\sum_a\xi_a$, cf. also
\cite{Mo,WeHa}. In this case we are able to write
\begin{equation}
  \label{eq:rs}
  c(\vec\xi) = \int dh\ P(h) \frac{\exp(h\sum_a\xi_a)}{(1+e^h)^n}
\end{equation}
with $P(h)$ being a probability distribution to guarantee the
normalization of $c(\vec\xi)$. The physical interpretation of $P(h)$
is simple: Take any vertex $i$, then its average local occupation
number $\langle x_i\rangle_\mu$ in the presence of the chemical
potential $\mu$ can be written as $e^{h_i}/(1+e^{h_i})$ using an
effective chemical potential $h_i$ accounting for all interactions on
$i$. $P(h)$ can now be constructed as the histogram of these effective
chemical potentials.

Plugging this ansatz into equations (\ref{eq:freeener}) and
(\ref{eq:sp}), the replica number $n$ appears as a mere parameter, and
the limit $n\to 0$ can be calculated. Details of this calculation are
given in appendix \ref{app:rs}. The saddle-point equation
(\ref{eq:sp}) now reads
\begin{equation}
  \label{eq:saddle}
  \int dh\ P(h)e^{h y} = \exp\left\{-c+\mu y +c\int dh\ P(h) 
  (1+e^h)^{-y} \right\}
\end{equation}
and has to be valid for arbitrary $y$. According to equations
(\ref{eq:entropy},\ref{eq:vcentr}) we find the entropy density of
vertex covers using a fraction $x=1-\int dh\ P(h)/(1+e^{-h})$ of 
vertices
\begin{eqnarray}
  \label{eq:srs}
  s_{VC}(x) &=& \int \frac{dh\ dk}{2\pi} e^{ikh} P_{FT}(k) [1-\ln
  P_{FT}(k) ] \ \ln (1+e^{h}) \nonumber\\ 
  && +\frac{c}{2} \int dh_1\ dh_2 P(h_1)
  P(h_2) \ln\left( 1-\frac{1}{(1+e^{-h_1})(1+e^{-h_2})} \right)
\end{eqnarray}
where
\begin{equation}
  \label{eq:fourier}
  P_{FT}(k) := \int dh\ e^{-ikh}\ P(h)
\end{equation}
denotes the Fourier transform of $P(h)$.

\subsection{Size of minimal vertex covers}\label{sec:xc}

It is however to complicated to directly solve (\ref{eq:saddle}) for
arbitrary $\mu$. But we are interested in the properties of minimal
vertex covers, which, according to section \ref{sec:latticegas} can be
described by the limit $\mu\to\infty$ of infinitely large chemical
potential. In this case, we also expect the effective chemical
potentials $h$ to scale as $z\mu$, with $z$ being a random variable
with finite mean and variance. The rescaled probability distribution
is denoted by $\tilde{P}(z)$. Please note that a negative $z$ now
corresponds to vertices having always $x_i=0$, whereas positive $z$
indicate vertices with fixed $x_i=1$. All vertices being occupied in
some ground-states and empty in others are collected in $z=0$. This
picture has to be refined for the calculation of the vertex-cover
entropy in section \ref{sec:sc}: There also contributions of
$O(\mu^0)$ have to be taken into account. For the present purpose the
dominant terms are however sufficient. To obtain a well-defined limit
$\mu\to\infty$ of eq. (\ref{eq:saddle}), we also have to rescale $y$ 
by $k:=\mu y$. Thus eq. (\ref{eq:saddle}) becomes
\begin{equation}
  \label{eq:saddle2}
  \int dz\ \tilde{P}(z) e^{zk} = \exp\left\{-c+ z
    +c\left(\int_{-\infty}^{+0}dz\ \tilde{P}(z)+\int_{+0}^\infty
    dz\  \tilde{P}(z) e^{-zk} \right)\right\}\ .
\end{equation}
The interpretation of this equation in terms of the cavity approach,
see \cite{MePaVi}, becomes evident if we Fourier-transform it and 
develop the last part of the exponential on the right-hand side,
\begin{eqnarray}
  \label{eq:saddle3}
  \tilde{P}(z) &=& \sum_{d=0}^\infty e^{-c} \frac{c^d}{d!} \left[
  \delta(\cdot+1)*\tilde{P}_-^{* d}(\cdot) \right](z)\ ,
\nonumber\\ 
  \tilde{P}_-(z) &=& \delta(z)\int_{-\infty}^{+0}dz\ \tilde{P}(z)
  +\Theta(-z)\tilde{P}(-z)\ .
\end{eqnarray}
$*$ denotes the convolution product.  This equation describes the
effective chemical potentials of a vertex which is the linear
superposition of the exterior chemical potential $1\cdot\mu$ and the
contributions of all neighbors. The contribution $\tilde{P}_-(z)$ of a
neighbor depends on the effective potentials the neighbors would have
without the presence of the central vertex, for details on this cavity
interpretation see \cite{MePaVi}, and reflects the hard-sphere
condition. A neighbor with positive potential would be occupied,
$x_i=1$, and thus forces a negative field for the central
term. Neighbors with non-negative chemical potential do not impose
anything as they would have $x_i=0$.  At the end, the resulting
distribution is averaged over the connectivity distribution of random
graphs.

This saddle-point equation has the simple solution
\begin{equation}
  \label{eq:pofz}
  \tilde{P}(z) = \sum_{l=-1}^\infty \frac{W(c)^{l+2}}{(l+1)!\ c} 
  \delta(z+l)
\end{equation}
where the Lambert-W function $W(c)$ is defined as the real solution
of
\begin{equation}
  \label{eq:w}
  c = W(c)\ e^{W(c)}\ .
\end{equation}
We already mentioned that vertices having negative fields are frozen
to $x_i=0$, vertices with positive fields to $x_i=1$. At the present
level, we are however not able to calculate the average magnetization
of the vertices belonging to $z=0$. This will be done in the next
section. Here we only use the result: Half of $(z=0)$-vertices are
occupied, half are empty. We therefore find an average occupation
density of hard spheres,
\begin{equation}
  \label{eq:density}
  \nu(\mu\to\infty)
  = \frac{1}{N}\langle\sum_i x_i \rangle_{\mu\to\infty} 
  = \frac{2W(c)+W(c)^2}{2c}\ ,
\end{equation}
which translates to a minimal vertex-cover size given by
\begin{equation}
  \label{eq:xc}
  x_c(c) = 1- \frac{2W(c)+W(c)^2}{2c}\ .
\end{equation}
In figure \ref{figXCC}, this result is compared to numerical
simulations. Extremely good coincidence is found for small
connectivities $c$. For non-percolating graphs, {\it i.e.} for $c<1$,
our result was recently proven to be exact \cite{BaGo}.

Systematic deviations show up later. For large $c$, eq. (\ref{eq:xc})
even violates the bounds given in section \ref{sec:bound} and the
exactly known asymptotics (\ref{asympt}). As we will see later, this
can be explained within our approach: Replica symmetry breaks down at
$c=e\simeq 2.718$, see the following sections. Up to this value
however, we expect the replica-symmetric result to be exact. This is
astonishing, as the solution does not show any particular signature of
the percolation transition of the underlying random graph at $c=1$.

\subsection{The backbone}\label{sec:bb}

The distribution $\tilde{P}(z)$ contains much more statistical
information on minimal vertex covers than simply its size. One
important effect is a partial freezing which can be observed: There
are vertices, which are always uncovered ($x_i=1$) in all minimal
vertex covers, others are always covered ($x_i=0$). We call these
spins uncovered (resp. covered) {\it backbone}. The fractions of 
vertices belonging to these two backbone types are given by
\begin{eqnarray}
  \label{eq:bb}
  b_{uncov}(c) &=& \frac{W(c)}{c} \nonumber\\
  b_{cov}(c) &=& 1-\frac{W(c)+W(c)^2}{c} \ .
\end{eqnarray}
The remaining $W(c)^2N/c$ vertices are unfrozen, their covering state 
changes from ground state to ground state. These different freezing
properties can already be seen in simple finite graphs: A graph
consisting of two connected vertices has two minimal vertex covers,
and the state of the two vertices is not uniquely determined. They
thus do not belong to the backbone. The situation is different for 
graphs with three vertices and two edges. The central vertex is
covered in the unique minimal vertex cover, thus belonging to the
covered backbone. The other two vertices form the negative backbone.

Let us now investigate the influence of the close environment of a
vertex on its behavior, more precisely the influence of its
connectivity. The total connectivity distribution is given by the
Poisson law (\ref{eq:Poisson}), but we can distinguish three distinct
contributions:
\begin{itemize}
\item The joint probability $P(d,\langle x\rangle=1)$ that a vertex
  has connectivity $d$ and belongs to the uncovered backbone of all
  minimal vertex covers.
\item $P(d,\langle x\rangle=0)$ gives the probability that a vertex 
  has connectivity $d$ and belongs to the covered backbone of all
  minimal vertex covers.
\item The remaining vertices are not in the backbone, thus
  being described by $P(d,0<\langle x\rangle<1)$.
\end{itemize}
These quantities can be easily computed from $\tilde{P}(z)$: according
to the interpretation of the self-consistent equation (\ref{eq:saddle3})
we can calculate the effective-field distribution for a vertex
of connectivity $d$ which, in average, has typical neighbors:
\begin{equation}
  \label{P_d}
  \tilde{P}_d (z) =  
  \left[\delta(\cdot+1)*\tilde{P_-}^{* d}(\cdot)\right](z)
\end{equation}
where $\tilde{P}_-(z)$ is exactly the quantity given in 
(\ref{eq:saddle3}). Plugging our solution (\ref{eq:pofz}) into
this equation, we find
\begin{eqnarray}
P(d,\langle x\rangle=1) = \tilde{P}_d (z>0) Po_{c}(d) &=&
    e^{-c} \frac{[c-W(c)]^d}{d!} \nonumber\\
P(d,0<\langle x\rangle<1) = \tilde{P}_d (z=0) Po_{c}(d) &=&
    e^{-c} \frac{W(c)[c-W(c)]^{d-1}}{(d-1)!} \\
P(d,\langle x\rangle=0) = \tilde{P}_d (z<0) Po_{c}(d) &=&
    e^{-c} \frac{[c+(d-1)W(c)][c-W(c)]^{d-1}}{(d-1)!}\nonumber
\end{eqnarray}
The results for $c=2$ are displayed in Fig. \ref{figConn} along with
numerical data for $N=17,35,70$. Please note that the numerical
results seem to converge towards the analytical one, thus showing an
excellent coincidence of both approaches. The curves are easily
understood: A vertex of connectivity 0 has no neighbors. Therefore, it
does not appear in any optimum cover and we obtain $P(0,\langle
x\rangle=1)=Po_c(0)$, $P(0,\langle x\rangle<1)=0$.  With increasing
connectivity the probability that a vertex is covered increases, thus
the contribution of $P(d,0<\langle x\rangle\leq1)$ to
$\mbox{Po}_{c}(d)$ increases as well. For large connectivities it is
very probable that a vertex belongs to all VCs but even a finite
fraction of vertices with $\langle x\rangle=1$ remains.  These results
justify {\it a posteriori} the application of a greedy heuristic
within the algorithm: Vertices having large connectivity are at first
included into the cover set.

\subsection{Entropy of minimal vertex covers}\label{sec:sc}

In order to calculate the entropy of minimal vertex covers, we have to
go beyond the leading terms in the effective chemical potentials. If
we consider {\it e.g.} the non-backbone spins, the order $\mu$ of the
effective fields is vanishing, but the order $\mu^0$ determines the
actual average occupation. We thus have to decompose the effective
potentials according to $h=\mu z + \tilde{z}$, and write the order
parameter as
\begin{equation}
  \label{eq:ptwovar}
  c(\vec\xi) = \int dz\ d\tilde{z}\ P(z,\tilde{z}) 
  \frac{\exp((\mu z+\tilde{z}) \sum_a\xi^a )}{(1+e^{\mu z+\tilde{z}})^n}
\end{equation}
where both $z$ and $\tilde{z}$ stay finite in the limit
$\mu\to\infty$. In this sense, we have $\int d\tilde{z}\
P(z,\tilde{z}) = \tilde{P}(z)$, the effective distribution calculated
in (\ref{eq:pofz}). We thus write (for $\mu\to\infty$)
\begin{eqnarray}
  \label{eq:ansatz}
  P(z,\tilde{z}) &=& \sum_{l=-1}^\infty p_l
  \delta(z+l) \rho^{(l)}(\tilde{z})\nonumber\\
  p_l &=& \frac{W(c)^{l+2}}{(l+1)!\ c}
\end{eqnarray}
where $\rho^{(l)}(\tilde{z})$ describes the still unknown sub-dominant
contributions to the effective potential $-\mu l$.  Plugging ansatz
(\ref{eq:ansatz}) into equation (\ref{eq:freeener}) for the grand
partition function, we see that the dominating part in $\ln \Xi$ is
linear in $\mu$, but vanishes finally once the saddle-point condition
is used. As is shown in appendix \ref{app:srs}, the term of $O(\mu^0)$
can be calculated and leads to the entropy of minimal vertex covers,
\begin{eqnarray}
  \label{eq:sminvc}
  s_{VC}(x_c(c)) &=& \int \frac{dz\ dk}{2\pi} \int \frac{d\tilde{z}\
    d\tilde{p}}{2\pi}\ e^{izk+i\tilde{z}\tilde{k}}\ P_{FT}(k,\tilde{k})
  \ [1-\ln P_{FT}(k,\tilde{k})] \ \Phi(z,\tilde{z})\nonumber\\
&& +\frac{c}{2}\ p_{-1}^2 \int d\tilde{z}_1\ d\tilde{z}_2\
  \rho^{(-1)}(\tilde{z}_1)\ \rho^{(-1)}(\tilde{z}_2)\ \ln(e^{-\tilde{z}_1}+
  e^{-\tilde{z}_1}) \nonumber\\
&& +c p_0 p_1 \int d\tilde{z}\ \rho^{(0)}(\tilde{z})\ \ln\left(1-
  \frac{1}{1+e^{-\tilde{z}}} \right) \nonumber\\
&& +\frac{c}{2} p_{0}^2 \int d\tilde{z}_1\ d\tilde{z}_2\
  \rho^{(0)}(\tilde{z}_1)\ \rho^{(0)}(\tilde{z}_2)\ \ln\left(1-
  \frac{1}{(1+e^{-\tilde{z}_1})(1+e^{-\tilde{z}_2})} \right)
\end{eqnarray}
where $P_{FT}(k,\tilde{k})$ signifies the two-dimensional Fourier 
transform of distribution $P(z,\tilde{z})$, and
\begin{equation}
  \label{eq:Phi}
  \Phi(z,\tilde{z}) = \left\{
    \begin{array}{lll}
    0 & \mbox{if} & z<0\\
    \ln (1+e^{\tilde{z}}) &\mbox{if}& z=0\\
    \tilde{z} &\mbox{if}& z>0\\
    \end{array}
\right.
\end{equation}
is the $O(\mu^0)$-term in $\ln(1+e^{\mu z+\tilde{z}})$.  The
corresponding saddle-point equations for the densities
$\rho^{(l)}(\tilde{z})$, which are also calculated in appendix
\ref{app:srs}, read
\begin{eqnarray}
  \label{eq:sp2}
  \rho^{(-1)}_{FT}(\tilde{k}) &=& \exp\left\{ -cp_0+cp_0 \int
    d\tilde{z}\ \rho^{(0)}(\tilde{z})\ (1+e^{\tilde{z}})^{-i\tilde{k}}  
    \right\}\nonumber\\
 \rho^{(l)}_{FT}(\tilde{k}) &=& \rho^{(-1)}_{FT}(\tilde{k})\  
    \rho^{(-1)}_{FT}(-\tilde{k})^{l+1}\ . 
\end{eqnarray}
We now easily see that $\rho^{(0)}$ is an even distribution, and
exactly half of the non-backbone vertices are covered in all minimal
vertex covers. Within the region of validity of the replica-symmetric
ansatz, this result is also verified numerically, see figure
\ref{fig:magnonbb}. Our argument for deriving equation (\ref{eq:xc})
for the minimal vertex-cover size is thus completed.  Using these
saddle-point equations, we can eliminate all but one $\rho^{(l)}$ in
the entropy (\ref{eq:sminvc}). After a lengthy calculation which is
again delegated to the appendix, we finally get
\begin{eqnarray}
  \label{eq:smin}
  s_{VC}(x_c(c)) &=& \frac{p_0}{2} \int \frac{d\tilde{z}\ 
  d\tilde{k}}{2\pi}
  e^{i\tilde{z}\tilde{k}} \rho^{(0)}_{FT}(\tilde{k}) [1-\ln
  \rho^{(0)}_{FT}(\tilde{k}) ] \ \ln (1+e^{\tilde{z}}) \nonumber\\ 
  && +\frac{cp_0^2}{2} \int d\tilde{z}_1\ d\tilde{z}_2\ 
  \rho^{(0)}(\tilde{z}_1)\ \rho^{(0)}(\tilde{z}_2)\  
  \ln\left( 1-\frac{1}{(1+e^{-\tilde{z}_1})(1+e^{-\tilde{z}_2})} \right)
\end{eqnarray}
which formally equals the expression (\ref{eq:srs}) for the
vertex-cover entropy for finite chemical potential $\mu$, with $c$
replaced by $2cp_0=2W(c)^2$. The main difference is however the
self-consistent equation 
\begin{equation}
  \label{eq:rho0}
  \rho^{(0)}_{FT}(\tilde{k}) = \exp\left\{ -2cp_0+cp_0 \int
    d\tilde{z} \rho^{(0)}(\tilde{z}) \left[
      (1+e^{\tilde{z}})^{i\tilde{k}} + (1+e^{\tilde{z}})^{-i\tilde{k}}
    \right]\right\}
\end{equation}
which can be obtained from (\ref{eq:smin}) by optimization with
respect to all {\it even} distributions $\rho^{(0)}(\tilde{z})$.
We are however not able to solve the last equation, and are therefore
restricted to variational approaches similar to \cite{BiMoWe}. A first
upper estimate would be
\begin{equation}
  \label{eq:supper}
  s_{VC}(x_c(c)) \simeq \frac{p_0}{2} \ln 2 +
  \frac{cp_0^2}{2}\ln\frac{3}{4}
\end{equation}
resulting from $\rho^{(0)}_{var}=\delta(\tilde{z})$. This results can
be slightly improved by using a Gaussian variational ansatz for
$\rho^{(0)}$, but the difference is only up to about $1\%$. For a
comparison with numerical results see figure \ref{figSCC}.

From equation (\ref{eq:rho0}) we are also able to read off
analytically some limitations of the replica-symmetric solution
(\ref{eq:pofz}). By developing (\ref{eq:rho0}) two second order in
$\tilde{k}$, we find
\begin{eqnarray}
  \label{eq:var}
 \Delta^2 &:=& \int d\tilde{z} \rho^{(0)}(\tilde{z}) \tilde{z}^2 
   \nonumber\\
  &=& 2cp_0 \int_0^\infty d\tilde{z} \rho^{(0)}(\tilde{z}) \left\{
  \ln(1+e^{-\tilde{z}})^2+\ln(1+e^{\tilde{z}})^2 \right\} \ .  
\end{eqnarray}
Rescaling $\tilde{z}=\Delta z$, we find 
\begin{equation}
  \label{eq:rescale}
  1 = \frac{2cp_0}{\Delta^2} \int_0^\infty dz\ \tilde{\rho}(z) \left\{
  \ln(1+e^{-\Delta z})^2+\ln(1+e^{\Delta z})^2 \right\}
\end{equation}
with $\tilde{\rho}(z)=\rho^{(0)}(z/\Delta)/\Delta$ being of unit
variance. For any $\tilde{\rho}$, the right-hand side is an
monotonously decreasing function of $\Delta$ ranging from $+\infty$
for $\Delta_0=0$ to $cp_0=W(c)^2$ for $\Delta\to\infty$. Identity
(\ref{eq:rescale}) can thus be satisfied if and only if $W(c)^2<1$,
which is valid for $c<e\simeq 2.718$. We thus have to conclude that
our replica-symmetric solution (\ref{eq:pofz}) becomes inconsistent
beyond average connectivity $e$, which is again in perfect agreement
with the systematic deviations of numerical data beyond this point,
cf. figures \ref{figXCC} and \ref{figBCC}. Note however, that this
point is far beyond the percolation threshold of the random
graph. After the next subsection we will come back to this point, and
consider more involved replica-symmetric and one-step broken saddle
points.

\subsection{The structure of the non-backbone subgraph}\label{sec:nonbb}

Before doing this, we will complete the discussion of the structure of
minimal vertex covers in the region $0<c<e$, where the described
solution is expected to be exact. In this subsection we will
concentrate on the structure of the non-backbone subgraph, {\it i.e.}
the subgraph composed of all vertices which are not in the backbone,
and all edges from $E$ connecting these vertices.

The first intuition on the structure of the non-backbone component can
be drawn from the saddle-point equation (\ref{eq:saddle3}) for the
distribution of effective chemical potentials. We consider an
arbitrary vertex, and call the graph reduced which is obtained from
the original graph by deleting the considered vertex and all its
incident edges. According to the cavity interpretation of
(\ref{eq:saddle3}), the vertex is not in the backbone iff exactly one
of its neighbors would be in the uncovered backbone of the reduced
graph. The meaning of this becomes evident if we consider the
non-backbone graphs in figure \ref{fig:nonbb}: Take {\it e.g.} the
graph consisting of four vertices and three edges. All its vertices
belong to the non-backbone. Deleting a boundary
vertex, the reduced graph becomes backbone, in particular the neighbor
of the boundary vertex belongs to the uncovered backbone. Deleting
instead a central vertex, the reduced subgraph becomes disconnected
into an isolated vertex, being uncovered backbone, and a connected
vertex pair, being non-backbone.

Iterating this argument, we conclude that the non-backbone can be
partitioned into $p_0N/2$ pairs of vertices, every pair containing an
edge and being eventually connected to other pairs or to covered
backbone vertices. The supplementary edges connecting different
pairs are conjectured to be drawn randomly and independently with 
the original probability $c/N$ between any non-backbone vertices.

Even if we are not able to prove this conjecture, we may give strong
arguments to support it:
\begin{itemize}
\item Looking at the non-backbone subgraphs of small tree-like graphs,
  the predicted structure is found. A cluster expansion up to
  connected clusters of four vertex pairs provides lower and upper
  bounds for the entropy which are in good agreement with numerical
  findings ({\it e.g.} in the first four non-zero digits for $c=0.1$).
\item We can apply the statistical-mechanics formalism to a restricted
  random graph ensemble having exactly the properties described above.
  This directly leads to expressions (\ref{eq:smin}) and (\ref{eq:rho0})
  for the entropy and the effective-potential distribution.
\item The proposed structure results in an even distribution of
  effective potentials for connectivities $c<e$, whereas the average
  occupation density is expected to exceed 1/2 for $c>e$. This is
  verified numerically, see figure \ref{fig:magnonbb}.
\item The average connectivity of a vertex pair to other vertex pairs
  in the restricted ensemble is $2cp_0=2W(c)^2$, the percolation
  threshold would therefore be at $W(c)=1/\sqrt{2}$, {\it i.e.} at
  $c=\exp(1/\sqrt{2})/\sqrt{2}\simeq 1.434$. We have checked this
  numerically by calculating the fraction of non-backbone vertices in
  the largest connected component of the non-backbone subgraph, see
  Fig. \ref{figFracLargest}. This quantity clearly extrapolates to 0
  for connectivities below the percolation point, and saturates at a
  finite value for larger connectivities. The reason why this
  transition is shifted to higher connectivity compared to graph
  percolation, becomes obvious by considering the action of covered
  backbone vertices. They ``cut'' the graph into smaller
  pieces. Please remember also that high-connectivity vertices are
  more frequently found in the covered backbone, making this cutting
  mechanism more effective.
\item We should add the remark that we have performed a similar 
  numerical study for the backbone subgraph. We found that
 the percolation threshold of the backbone subgraph is
  identical to the percolation threshold $c=1.0$ of the whole graph.
\end{itemize}

The percolation however does not bother the validity of the
replica-symmetric result, which is valid even for percolated
non-backbone subgraphs as long as $c<e$. The proposed structure also
allows for a very simple interpretation of approximation
(\ref{eq:supper}) of the entropy of minimal VCs: An isolated pair
contributes an entropy $\ln 2$ as it has two possible minimal VCs,
thus explaining the first term in (\ref{eq:supper}).
This entropy is decreased by the insertion of supplementary edges.
The simplest structure are chains of four vertices, every one having
only three minimal VCs -- leading directly to the second term in
(\ref{eq:supper}) as two pairs are connected with probability $4c/N$.
More complicated non-backbone graphs lead to corrections, and may be 
included by a non-trivial $\rho^{(0)}$.

\subsection{Unphysical replica-symmetric saddle 
            points}\label{sec:multipeak}

We have seen, that solution (\ref{eq:pofz}) of the replica-symmetric
saddle-point equation (\ref{eq:saddle3}) is correct only up to average
connectivity $c=e$. Before searching for replica-symmetry-broken
saddle points, we should however exclude the existence of further
replica-symmetric saddle points. We therefore consider again equation
(\ref{eq:saddle3}). It is consistent with any ansatz
\begin{equation}
  \label{eq:multipeak}
  \tilde{P}_m(z) = \sum_{l=-m}^\infty p_l^{(m)} \delta\left(
  z+\frac{l}{m}\right)\ .
\end{equation}
One can easily write down the self-consistent conditions for the
probabilities $p_l^{(m)}$, and find out that, for $m>1$, these have 
non-trivial solutions with positive $p_l^{(m)}$s only for $c>e$. 
We will show this explicitly only for $m=2$. The saddle-point 
equations read
\begin{eqnarray}
  \label{eq:twopeak}
  p_{-2} &=& \exp\{-c(p_{-2}+p_{-1})\} \nonumber\\
  p_{-1} &=& \exp\{-c(p_{-2}+p_{-1})\} cp_{-1} \nonumber\\
  p_0    &=& \exp\{-c(p_{-2}+p_{-1})\} (cp_{-2}+\frac{c}{2}p_{-1}^2)
   \nonumber\\      &\cdots&
\end{eqnarray}
The only solution with non-zero $p_{-1}$ is 
\begin{eqnarray}
  \label{eq:twosol}
  p_{-2} &=& \frac{1}{c} \nonumber\\
  p_{-1} &=& \frac{\ln c-1}{c} \nonumber\\
  p_{0}  &=& \frac{2+(\ln c-1)^2}{2c} \nonumber\\
         &\cdots&
\end{eqnarray}
$p_{-1}$ is obviously positive only if $c>e$.

The corresponding threshold $x_c(c)$ would be larger than the old one 
resulting from $m=1$, which is correct compared to the systematic
deviation of the numerical data. The multi-peak solutions
(\ref{eq:multipeak}) are however unphysical due to the existence of
effective potentials of {\it e.g.} the value $\mu/m$. This positive
potential would force the corresponding vertex to be in the uncovered
backbone for large $\mu$, but the only physical mechanism for this 
is given by the global chemical potential $\mu$. The influence of
neighbors results only in negative or zero potentials. Positive
fractions of  $\mu$ are consequently unphysical.

We can however interpret multi-peak solutions as a kind of hidden
replica-symmetry breaking. This will become clear in the following
section. 


\section{The simplest one-step replica symmetry broken 
        solution}\label{sec:1rsb}

This section is dedicated to the appearance of replica-symmetry breaking
(RSB) in VC. Despite several efforts, the question of how to handle
RSB in finite-connectivity systems is still open. Most attempts
\cite{DoMo,MoDo,WoSh,GoLa} try to apply the first step of Parisi's RSB
scheme \cite{MePaVi} which however was originally developped for
infinite-connectivity spin glasses. Due to the more complex structure
of the order parameter a complete solution is however still missing.
Recently, based on the connection to combinatorial optimization, the
interest in this question was renewed \cite{Mo}, and some promising
approximation schemes \cite{BiMoWe,MePa2} have been developed. Here
we closely follow the approach proposed in \cite{Mo} which allows to
construct a simple one-step RSB solution.

In case of one-step RSB, the full permutation symmetry of the
order parameter corresponding to the equivalence of all $n$ replicas
breaks down. According to Parisi's scheme, the replicas can be grouped
into $n/m$ blocks of equal size $m$, where the symmetry is now
restricted to permutations of replicas within every block, or to
permutations of full blocks. We therefore introduce a new numbering of
replicas by index pairs $(a,\alpha)$, with $a=1,...,n/m$ denoting the
block number, and $\alpha=1,...,m$ counting the replicas within block
$a$. Due to the described symmetry, the order parameters
$c(\vec\xi)$  thus depend on $\vec\xi$ only via the block
quantities $s^a=\sum_{\alpha=1}^m\xi^{a\alpha}$, or even more 
precisely, on the number of blocks having $s^a=ym$, which can be
described by
\begin{equation}
  \label{eq:nu}
  \nu(y)=\sum_{a=1}^{n/m} \delta(y-s^a/m)\ .
\end{equation}
$y$ stands for the average occupation number of a block and ranges 
from 0 to 1. Its discrete nature present for natural $n$ vanishes in 
the analytical continuation needed for the replica limit $n\to 0$.

Following the cavity-like argumentation of Monasson \cite{Mo}, the
order parameter can be expressed as
\begin{eqnarray}
  \label{eq:1rsb}
  c(\vec\xi)&=&\int {\cal{D}}\rho\ {\cal{P}}[\rho]\prod_{a=1}^{n/m}
  \int dh\ \rho(h)\ \frac{e^{hs^a}}{(1+e^h)^m}\nonumber\\
  &=& \int {\cal{D}}\rho\ {\cal{P}}[\rho]\ \exp\left\{
  \int_0^1 dy\ \nu(y) \ln  \left[ 
  \int  dh\ \rho(h)\ \frac{e^{hmy}}{(1+e^h)^m} \right] \right\}
  \nonumber\\ &=:& c[\nu]
\end{eqnarray}
where ${\cal{P}}[\rho]$ is a histogram of the local distributions
$\rho_i(h)$, which themselves are histograms of local effective 
potentials over all thermodynamically relevant pure states, see
\cite{MePaVi} for a detailed discussion of this interpretation.
In the second line, the analytic continuation in $n$ has already been
made, $m$ is now considered as usually as a parameter in the interval 
$[0,1]$ which has to be optimized in the saddle-point solution. The
only requirement to $\nu(y)$ is, that
\begin{equation}
  \label{eq:nuint}
  \int_0^1 dy\ \nu(y) = \frac{n}{m} \to 0
\end{equation}
vanishes in the replica limit $n\to 0$.

This ansatz can be plugged into the saddle-point equation (\ref{eq:sp})
\begin{equation}
  \label{eq:sprsb}
  c(\vec\xi) = \exp\left\{ -c +\mu\sum_{a\alpha}\xi^{a\alpha}
   +c\sum_{\vec\zeta} c(\vec\zeta) 
  \prod_{a\alpha}(1-\xi^{a\alpha}\zeta^{a\alpha})  \right\}\ .
\end{equation}
Proceeding term by term on the right-hand side, we find
\begin{equation}
  \label{eq:yav}
  \sum_{a\alpha}\xi^{a\alpha}=\sum_a s^a=m\int_0^1dy\ v(y)\ y
\end{equation}
and
\begin{eqnarray}
  \label{eq:rhsrsb}
  \sum_{\vec\zeta} c(\vec\zeta) 
  \prod_{a\alpha}(1-\xi^{a\alpha}\zeta^{a\alpha})&=&\int
  {\cal{D}}\rho\ {\cal{P}}[\rho]\prod_{a=1}^{n/m} \int dh\ \rho(h)\ 
  \prod_{\alpha=1}^m\left[\sum_{\zeta=0,1}
  \frac{e^{h\zeta}(1-\xi^{a\alpha}\zeta)}{1+e^h}\right]\nonumber\\
  &=& \int{\cal{D}}\rho\ {\cal{P}}[\rho]\prod_{a=1}^{n/m} \int dh\ 
  \rho(h)\ \left[ \frac{1}{1+e^h}\right]^{s^a}\nonumber\\
  &=&  \int{\cal{D}}\rho\ {\cal{P}}[\rho]\ \exp\left\{
  \int_0^1dy\ \nu(y)\ \ln\left[ \int dh\ \rho(h)\ (1+e^h)^{-my}
  \right] \right\} \nonumber
\end{eqnarray}
Plugging this results together with equations (\ref{eq:1rsb}) and
(\ref{eq:yav}) into (\ref{eq:sprsb}), we obtain for $n=0$ a closed 
equation for ${\cal{P}}[\rho]$ which has to be fulfilled for every
$\nu(y)$ satisfying condition (\ref{eq:nuint}).

This saddle-point equation is still valid for any chemical
potential. In the limit of minimal vertex covers, {\it i.e.} for
$\mu\to\infty$, this equation simplifies again. For the $\rho(h)$ we
assume an ansatz similar to the replica-symmetric value for $P(h)$
in (\ref{eq:pofz})
\begin{equation}
  \label{eq:rhoofh}
  \rho(h) = \omega_\mu \sum_{l=l_-}^{l_+} \rho_l e^{\mu m |l|/2} 
  \delta(h+\mu l)
\end{equation}
where the support of $\rho$ is now restricted by $l$-values between
$(l_-,l_+)$ with $l_-\geq -1$. This $l$-intervall changes from
instance to instance drawn from ${\cal{P}}[\rho]$. The normalizing
prefactor $\omega_\mu$ becomes irrelevant for $n\to 0$ due to
condition (\ref{eq:nuint}).  The exponential factor is inspired by its
appearance in infinite-connectivity models, cf. \cite{Mo}. Please note
that the replica-symmetric case can be obtained by
$l_-=l_+$. Introducing weights ${\cal{P}}_{l_-,l_+}$ as the integrated
weight of all $\rho(h)$ having the same $(l_-,l_+)$, the order
parameter simplifies to
\begin{equation}
  \label{eq:oprsbminvc}
 \lim_{\mu\to\infty} c[\nu/\mu] = \sum_{-1\leq l_-\leq l_+}
  {\cal{P}}_{l_-,l_+} \exp(\nu_-l_- +\nu_+ l_+)  
\end{equation}
where $\nu_-= m \int_{1/2}^1 dy\ \nu(y)\ (1/2-y)$ and 
$\nu_+= m \int_0^{1/2} dy\ \nu(y)\ (1/2-y)$. Details of this 
calculation are delegated to appendix \ref{app:rsb}. Our saddle-point 
equation thus becomes
\begin{equation}
  \label{eq:rsbsaddle}
  \sum_{-1\leq l_-\leq l_+}{\cal{P}}_{l_-,l_+}\exp(\nu_-l_-+\nu_+l_+)
  = \exp\left\{-c-(\nu_-+\nu_+)+c{\cal{P}}_{-1,-1}e^{\nu_-+\nu_+}
  + c{\cal{P}}_{-1,0}e^{\nu_-} \right\}
\end{equation}
which has to be fulfilled for all $\nu_-,\nu_+$. Please note that the
$m$-dependence is completely disappeared \cite{note}. This equation
can be easily solved:
\begin{eqnarray}
  \label{eq:rsbsol1}
  {\cal{P}}_{-1,-1} &=& \frac{1}{c}\nonumber\\
  {\cal{P}}_{-1,0} &=& \frac{\ln(c) -1}{c}\nonumber\\
  {\cal{P}}_{l_-,l_+} &=& \frac{c^{l_++1}}{(l_-+1)!(l_+-l_-)!}
  {{\cal{P}}_{-1,-1}}^{l_-+2}{{\cal{P}}_{-1,0}}^{l_+-l_-}\nonumber\\
  &=& \frac{(\ln(c)-1)^{l_+-l_-}}{(l_-+1)!(l_+-l_-)!c}
\end{eqnarray}
Let us discuss this solution:
\begin{itemize}
\item At first we realize that ${\cal{P}}_{-1,0}$ is positive only for
  connectivities $c>e$. This is consistent with our previous finding
  that replica symmetry is restricted to smaller $c$.
\item Introducing $p_l$ as the sum over all ${\cal{P}}_{l_-,l_+}$
  having $l=l_-+l_+$, saddle-point equation (\ref{eq:rsbsol1}) reduces
  for $\nu_-=\nu_+$ to equations (\ref{eq:twopeak}) for the unphysical
  replica-symmetric saddle point showing half-integer valued effective
  potentials. This underlines the interpretation of these solutions as
  hidden-RSB solutions.
\item As we do not know the non-backbone magnetization in the RSB
  solution, we are only able to give lower and upper estimates for
  $x_c(c)$. The upper one,
  $x_c(c)<1-{\cal{P}}_{-1,-1}-{\cal{P}}_{-1,0}=1-\ln(c)/c$ coincides 
  with the rigorous upper bound of Gazmuri \cite{Ga}. The lower one 
  would be $x_c(c)>1-{\cal{P}}_{-1,-1}-{\cal{P}}_{-1,0}-
  {\cal{P}}_{-1,+1}-{\cal{P}}_{0,0}$. Having in mind the numerical
  result, that non-backbone effective potentials have a positive
  bias, we can however conclude $x_c(c)>1-{\cal{P}}_{-1,-1}-
  {\cal{P}}_{-1,0}-{\cal{P}}_{-1,+1}/2-{\cal{P}}_{0,0}/2$ which is
  slightly better than the replica-symmetric result. In figure
  \ref{figXCC} both results are nearly indistinguishable, so we have 
  omitted the RSB data from the figure.
\item Also the evaluation of the backbone-size is slightly subtle.
  In principle we would expect that backbone vertices have $\rho(h)$
  which are supported either only on positive or only on negative
  fields. This would result in
  \begin{eqnarray}
    \label{eq:bb1}
    b_{uncov}^{(1)}(c) &=& {\cal{P}}_{-1,-1} = \frac{1}{c}\nonumber\\
    b_{cov}^{(1)}(c) &=& \sum_{1\leq l_-\leq l_+} {\cal{P}}_{l_-,l_+}
    = 1-\frac{2}{e}\ .
  \end{eqnarray}
  Due to the existence of the exponential factors in ansatz
  (\ref{eq:rhoofh}) also ${\cal{P}}_{l_-,l_+}$ with $l_-\neq -l_+$
  lead to average occupation numbers zero and one, and thus contribute
  to the backbone:
  \begin{eqnarray}
    \label{eq:bb2}
    b_{uncov}^{(2)}(c) &=& {\cal{P}}_{-1,-1}+{\cal{P}}_{-1,0}
    = \frac{\ln c}{c}\nonumber\\
    b_{cov}^{(2)}(c) &=& \sum_{l_-\geq-1;l_+>|l_-|} {\cal{P}}_{l_-,l_+}
    \nonumber\\
    &=& 1- \frac{1}{c}\left(1+\ln c+\frac{(1-\ln c)^2}{4}\right)\ .
  \end{eqnarray}
  Both values do not coincide with numerical findings, see also figure
  \ref{figBCC}. Probably this could be cured by assuming $m\sim
  \mu^{-1}$, cf. \cite{note}, instead of $m\sim \mu^0$. This would
  remove the exponential dominance of fields of largest absolute value
  for $\mu\to\infty$. We could however construct no solution to this
  case. 
\end{itemize}
We may conclude that the presented one-step saddle point improves the
replica-symmetric findings for $x_c(c)$, but is still plagued by
certain problems. It remains an open question, if these problems can 
be cured already by including a different scaling of $m$, or if
finally more than one step of RSB is required.


\section{Conclusion and outlook}

In this paper, we have presented a detailed analysis of size and
structure of minimal vertex covers on random graphs. In particular, we
have calculated the size dependence of minimal VCs on the average
connectivity, and we ahve shown that those VCs are exponentially
numerous. Many statistical properties, as {\it e.g.} the partial
freezing into backbone and non-backbone vertices, could be
characterized. All our results are based on exact numerical
enumerations as well as replica calculations. We have found that
replica-symmetric results appear to be exact up to graph
connectivities $c=e\simeq 2.718$, whereas replica-symmetry breaking
has to be included for an understanding of higher-connectivity
graphs. This is however a complicated task: Even if there has been
some recent progress on the question of one-step replica-symmetry
breaking in finite connectivity systems based on various approximation
schemes \cite{Mo,BiMoWe,MePa2}, a definite technical approach is still
missing. Due to the simplicity of its replica-symmetric solution, as
compared e.g. to satisfiability problems \cite{MoZe}, vertex cover
could be a good model for further progress into this direction.

In our paper, we have only considered finite-connectivity random
graphs. These show however a very simple geometrical structure. They
are locally tree-like, and loops are of length $O(\ln N)$. It would
be interesting to considered therefore restricted graph ensembles
which include non-trivial local structures. The question of such
topological influences on the solution structure of combinatorial
optimization problems still remains an interesting open question, 
as also other studied problems include mainly locally tree-like
problems \cite{AI,TCS}. Restricted graph ensembles could therefore 
provide a possible starting point for further research.

A last comment concerns the interpretation of vertex covers as
packings of hard spheres on random lattices. We were able to describe
the maximally dense packings, which were found to show very
interesting properties due to the disorder present in the graph:
There where backbone sites having the same occupation state in all
densest packings, whereas others are found to be free in some
packings, occupied in others. This effect resembles the existence
of blocked and unblocked particles in real packings. With some
modifications, the hard-sphere lattice gas can therefore be understood
as a possible mean-field model of granular packings, compare also
\cite{granular5}. Work is in progress along these lines.

\section*{Acknowledgements}

The authors are deeply indebted to R. Monasson and R. Zecchina for 
many fruitful discussions. AKH acknowledges financial support by the 
DFG ({\em Deutsche Forschungsgemeinschaft}) under grant Zi209/6-1.


\appendix

\section{The replica-symmetric limit $n\to 0$}\label{app:rs}

Starting from  equations (\ref{eq:freeener}) and (\ref{eq:sp})
we will present the calculation of the replica limit $n\to 0$
under the replica-symmetric ansatz
\begin{equation}
  \label{eq:ans}
  c(\vec\xi) = \int dh P(h) \frac{\exp(h\sum_a\xi^a)}{(1+e^h)^n}
\end{equation}
The procedure is very similar to the one presented in \cite{Mo} for
Ising-spin-glass models.
We start with the grand partition function as given in
(\ref{eq:freeener}):
\begin{equation}
  \label{eq:lnxi}
  \lim_{N\to\infty}\frac{1}{N}\overline{\ln\Xi}
   = \lim_{n\to 0}\frac{1}{n}
    \left(-\sum_{\vec\xi} c(\vec\xi)\ln c(\vec\xi) -\frac{c}{2}
        +\mu \sum_{\vec\xi,a}c(\vec\xi)\xi^a +\frac{c}{2}
        \sum_{\vec\xi,\vec\zeta}c(\vec\xi)c(\vec\zeta)
        \prod_a(1-\xi^a\zeta^a) \right)
\end{equation}
where $c(\vec\xi)$ takes its saddle-point value.
At first, we consider the combinatorial entropy, and use again a replica
trick:
\begin{equation}
  \label{eq:a}
  \sum_{\vec\xi} c(\vec\xi)\ln c(\vec\xi) = \left[
  \frac{\partial}{\partial l} \sum_{\vec\xi} c(\vec\xi)^l
  \right]_{l=1}
\end{equation}
Assuming positive integer $l$ at the beginning, and plugging in the
replica-symmetric ansatz for $c(\vec\xi)$, we  write
\begin{eqnarray}
  \label{eq:b}
  \sum_{\vec\xi} c(\vec\xi)^l &=& \int dh_1\cdots dh_l P(h_1)\cdots
  P(h_l) \sum_{\vec\xi} \frac{\exp(\sum_{m=1}^l h_m \sum_a \xi^a)}{
    \prod_{m=1}^l (1+e^{h_l})^n} \nonumber\\
  &=&  \int dh_1\cdots dh_l P(h_1)\cdots P(h_l) 
  \left( \sum_{\xi=0,1} \frac{\exp(\sum_l h_l \xi)}{
    \prod_{m=1}^l (1+e^{h_l})} \right)^n \nonumber\\
  &=& 1 + n \int dh_1\cdots dh_l P(h_1)\cdots P(h_l) \ln(1+\exp\{\sum_m
    h_m\}) \nonumber\\
  &&  -nl \int dh\ P(h) \ln(1+e^h) + O(n^2) \nonumber
\end{eqnarray}
Introducing new variables $H_k = \sum_{m=1}^k h_m$, the last
expression becomes
\begin{eqnarray}
  \label{eq:c}
  \sum_{\vec\xi} c(\vec\xi)^l &=& 1 + n \int dH_1\cdots dH_l P(H_1)
  P(H_2-H_1) \cdots P(H_l-H_{l-1}) \ln(1+e^{H_l})\nonumber\\ 
  &&-nl \int dh\ P(h) \ln(1+e^h) + O(n^2) \nonumber\\
  &=& 1 + n \int dH_l \int \frac{dk}{2\pi} e^{iH_lk} P_{FT}(k)^l 
  \ln(1+e^{H_l})-nl \int dh\ P(h) \ln(1+e^h) + O(n^2) \nonumber
\end{eqnarray}
In the last step we have used the fact, that the $l$-fold convolution
of $P(h)$ with itself can be express as the Fourier-back
transformation of the $l$th power of its Fourier transform $P_{FT}$.
Now the differentiation with respect to $l$ can be carried out, and 
according to (\ref{eq:a}) we find
\begin{equation}
  \label{eq:d}
  \sum_{\vec\xi} c(\vec\xi) \ln c(\vec\xi) = n \int \frac{dh\
    dk}{2\pi} e^{ihk} P_{FT}(k) \left[-1+\ln P_{FT}(k) \right] 
   \ln (1+e^h) \ .
\end{equation}
The other terms in (\ref{eq:lnxi}) can be evaluated directly,
\begin{eqnarray}
  \label{eq:e}
  \sum_{\vec\xi, a} c(\vec\xi)\ \xi^a &=& n \int dh\ P(h)\ \frac{e^h
    (1+e^h)^{n-1}}{(1+e^h)^n }\nonumber\\
  &=&  n \int dh\ P(h)\ \frac{1}{(1+e^{-h})}\nonumber\\
 \sum_{\vec\xi,\vec\zeta}c(\vec\xi)c(\vec\zeta)
 \prod_a(1-\xi^a\zeta^a) &=& \int dh_1dh_2\ P(h_1)\ P(h_2) 
 \sum_{\vec\xi,\vec\zeta} \prod_{a=1}^n
 \frac{(1-\xi^a\zeta^a)\exp(h_1\xi^a+h_2\zeta^a)}{(1+e^{h_1}) 
     (1+e^{h_2})}\nonumber\\
 &=& \int dh_1dh_2\ P(h_1)\ P(h_2) \left[ 1-\frac{e^{h_1+h_2}}{ 
   (1+e^{h_1}) (1+e^{h_2})}  \right]^n\nonumber\\
 &=& 1+ n \int dh_1dh_2\ P(h_1)\ P(h_2) \ln\left[ 1-
   \frac{e^{h_1+h_2}}{(1+e^{h_1}) (1+e^{h_2})}  \right] +O(n^2)
  \nonumber
\end{eqnarray}
Putting these results together, we find
\begin{eqnarray}
  \label{eq:f}
  \lim_{N\to\infty}\frac{1}{N} \overline{\ln\Xi} &=& \int \frac{dh\
    dk}{2\pi} e^{ihk} P_{FT}(k) \left[1-\ln P_{FT}(k) \right] 
   \ln (1+e^h) + \mu \int dh\ P(h)\ \frac{1}{(1+e^{-h})}\nonumber\\ 
  &&-\frac{c}{2} 
   \int dh_1dh_2\ P(h_1)\ P(h_2) \ln\left[ 1-
   \frac{e^{h_1+h_2}}{(1+e^{h_1}) (1+e^{h_2})}  \right]
\end{eqnarray}
which finally results in equation (\ref{eq:srs}) for the vertex-cover
entropy. 

For the saddle-point equation
\begin{equation}
  \label{eq:cofxi}
  c(\vec\xi) = \exp\left\{ -\lambda +\mu\sum_a\xi^a+c\sum_{\vec\zeta}
    c(\vec\zeta) \prod_a(1-\xi^a\zeta^a)  \right\}\ .
\end{equation}
we proceed analogously. Obviously both side depend on $\vec\xi$ only
via $y=\sum_a\xi^a$. The left-hand side thus simplifies for $n\to 0$:
\begin{equation}
  \label{eq:g}
  c(\vec\xi) \to_{n\to 0} \int dh\ P(h)\ e^{hy}\ ,
\end{equation}
whereas the right-hand side ($rhs$) gives
\begin{eqnarray}
  \label{eq:h}
  rhs &=& \exp\left\{ -\lambda +\mu y +c \int dh\ P(h) \left(
  \frac{1}{1+e^h}\right)^{y} \right\}
\end{eqnarray}
We now can determine the Lagrange multiplier from the normalization of
$P(h)$. For $y=0$, the left-hand side equals one, whereas the
right-hand side equals $\exp(-\lambda+c)$, which results directly in
$\lambda=c$, and thus in the replica-symmetric saddle-point equation
(\ref{eq:saddle}).  

The same saddle-point equation can of course be derived by varying
equation (\ref{eq:f}) directly with respect to $P(h)$. Note however
that the result given here is stronger: We have shown that the
original saddle-point equation for $c(\vec\xi)$ is closed under our
replica-symmetric ansatz, thus leading to a real saddle point of the
free energy. The second procedure would however be important if we
would use a variational ansatz which does not close the
$c(\vec\xi)$-equation. 

\section{Calculation of the entropy}\label{app:srs}

For calculating the entropy of minimal vertex covers, we start again
with equation (\ref{eq:freeener}) for the disorder-averaged grand
partition function, but now we plug in the refined ansatz
(\ref{eq:ptwovar}), {\it i.e.}
\begin{equation}
  \label{eq:Ba}
  c(\vec\xi) = \int dz\ d\tilde{z}\ P(z,\tilde{z}) 
  \frac{\exp((\mu z+\tilde{z}) \sum_a\xi^a )}{(1+e^{\mu z+\tilde{z}})^n}
\end{equation}
where $P(z,\tilde{z})$ is assumed to stay a well-behaved probability
distribution in the limit $\mu\to\infty$ of minimal vertex
covers. Consistency with the dominant behavior discussed in section
\ref{sec:xc} requires
\begin{equation}
  \label{eq:Ba1}
  \int d\tilde{z} P(z,\tilde{z}) = \sum_{l=-1} p_l \delta(z+l)
\end{equation}
with 
\begin{equation}
  \label{eq:Ba2}
  p_l = \frac{W(c)^{l+2}}{(l+1)!\ c}
\end{equation}
for $\mu\to\infty$, cf. (\ref{eq:pofz}). We therefore may write
\begin{equation}
  \label{eq:Ba3}
  P(z,\tilde{z}) = \sum_{l=-1} p_l \delta(z+l) \rho^{(l)}(\tilde{z})
\end{equation}
with probability distributions $\rho^{(l)}(\tilde{z})$ which still
have to be determined. By
plugging ansatz (\ref{eq:Ba}) into $\overline{\ln\Xi}$ as given in
(\ref{eq:freeener}) and following the same procedure as in the last
section, we find for finite $\mu$:
\begin{eqnarray}
  \label{eq:Bb}
    \lim_{N\to\infty}\frac{\overline{\ln\Xi}}{N} &=& \int \frac{dz\
    dk}{2\pi} \int \frac{d\tilde{z}\ d\tilde{k}}{2\pi}
  e^{izk+i\tilde{z}\tilde{k}}  P_{FT}(k,\tilde{k}) \left[1-\ln
    P_{FT}(k,\tilde{k}) \right] \ln (1+e^{\mu z+\tilde{z}})\nonumber\\ 
  && + \mu \int dz\ d\tilde{z}\ P(z,\tilde{z})\ \frac{1}{
   (1+e^{-\mu z-\tilde{z}})} \\ 
  && - \frac{c}{2} 
   \int dz_1dz_2d\tilde{z}_1d\tilde{z}_2 P(z_1,\tilde{z}_1)
   P(z_2,\tilde{z}_2) \ln\left[ 1-
   \frac{1}{(1+e^{-\mu z_1-\tilde{z}_1}) 
         (1+e^{-\mu z_2-\tilde{z}_2})}  \right] \nonumber
\end{eqnarray}
with $P_{FT}(k,\tilde{k})=\int dz\ d\tilde{z}\ P(z,\tilde{z})
\exp\{-izk-i\tilde{z}\tilde{k}\}$ being the 2d Fourier-transform of
$P(z,\tilde{z})$. For $\mu\to\infty$, the dominant behavior seems to
be of $O(\mu)$, but its coefficient has to vanish at the saddle point
as the entropy stays finite. This has been checked explicitly,
without presenting those details we therefore concentrate on the 
second term of $O(\mu^0)$ which will give the entropy of minimal vertex
covers,
\begin{equation}
  \label{eq:Bc}
  s_{VC}(x_c(c)) = \lim_{\mu\to\infty} \left(
    \lim_{N\to\infty}\frac{1}{N} \overline{\ln\Xi} 
    - \mu \int dz\ d\tilde{z}\ P(z,\tilde{z})\ \frac{1}{
   (1+e^{-\mu z-\tilde{z}})} \right)\ .
\end{equation}
Starting with the first term in (\ref{eq:Bb}), we have to leading
orders
\begin{eqnarray}
  \label{eq:Bd}
  \ln (1+e^{\mu z+\tilde{z}}) &\to& \mu z \Theta(z) +\Phi(z,\tilde{z}) 
  \nonumber\\
  \Phi(z,\tilde{z}) &=& \left\{
    \begin{array}{lll}
    0 & \mbox{if} & z<0\\
    \ln (1+e^{\tilde{z}}) &\mbox{if}& z=0\\
    \tilde{z} &\mbox{if}& z>0\\
    \end{array}
\right.
\end{eqnarray}
At the moment, replacing $\ln (1+e^{\mu z+\tilde{z}})$ by
$\Phi(z,\tilde{z})$ is all we can do in the first term without using
the saddle-point equations for the $\rho^{(l)}$s. The situation is
better for the last term in (\ref{eq:Bb}). Having in mind that $z$ can
take only integer values smaller or equal to $+1$, we use
\begin{equation}
  \label{eq:Be}
  \ln\left[ 1-
   \frac{1}{(1+e^{-\mu z_1-\tilde{z}_1}) 
         (1+e^{-\mu z_2-\tilde{z}_2})}  \right] \to \left\{
       \begin{array}{lll}
     -\mu + \ln(e^{-\tilde{z}_1}+e^{-\tilde{z}_2})  &\mbox{if}&
     z_1=z_2=1\\
     \ln  \left[ 1- \frac{1}{1+e^{-\tilde{z}_1}}  \right]
     &\mbox{if}& z_1=0,\ z_2=1\\
     \ln  \left[ 1- \frac{1}{1+e^{-\tilde{z}_2}}  \right]
     &\mbox{if}& z_1=1,\ z_2=0\\
     \ln\left[ 1-\frac{1}{(1+e^{-\tilde{z}_1}) (1+e^{-\tilde{z}_2})}  
     \right]      &\mbox{if}& z_1= z_2=0\\
     0   &\mbox{if}& z_1,z_2<0
       \end{array}
\right.
\end{equation}
where all terms are dropped which are exponentially small in $\mu$. 
Plugging this result into (\ref{eq:Bb}), we find
\begin{eqnarray}
  \label{eq:Bf}
  s_{VC}(x_c(c)) &=& \int \frac{dz\ dk}{2\pi} \int \frac{d\tilde{z}\
    d\tilde{p}}{2\pi}\ e^{izk+i\tilde{z}\tilde{k}}\ P_{FT}(k,\tilde{k})
  \ [1-\ln P_{FT}(k,\tilde{k})] \ \Phi(z,\tilde{z})\nonumber\\
&& +\frac{c}{2}\ p_{-1}^2 \int d\tilde{z}_1\ d\tilde{z}_2\
  \rho^{(-1)}(\tilde{z}_1)\ \rho^{(-1)}(\tilde{z}_2)\ \ln(e^{-\tilde{z}_1}+
  e^{-\tilde{z}_1}) \nonumber\\
&& +c p_0 p_1 \int d\tilde{z}\ \rho^{(0)}(\tilde{z})\ \ln\left(1-
  \frac{1}{1+e^{-\tilde{z}}} \right) \nonumber\\
&& +\frac{c}{2} p_{0}^2 \int d\tilde{z}_1\ d\tilde{z}_2\
  \rho^{(0)}(\tilde{z}_1)\ \rho^{(0)}(\tilde{z}_2)\ \ln\left(1-
  \frac{1}{(1+e^{-\tilde{z}_1})(1+e^{-\tilde{z}_2})} \right)
\end{eqnarray}
which is equation (\ref{eq:sminvc}).

We continue again with the derivation of the saddle-point equation,
again we start from the original equation for $c(\vec\xi)$ as given in
eq. (\ref{eq:sp}),
\begin{equation}
  \label{eq:Bg}
  c(\vec\xi) = \exp\left\{ -\lambda +\mu\sum_a\xi^a+c\sum_{\vec\zeta}
    c(\vec\zeta) \prod_a(1-\xi^a\zeta^a)  \right\}\ .
\end{equation}
Plugging in the replica-symmetric ansatz and continuing analogously to
the previous appendix for $n\to 0$, we find
\begin{equation}
  \label{eq:Bh}
  \int dz\ d\tilde{z} P(z,\tilde{z}) e^{\mu z k +\tilde{z}k} = 
  \exp\left\{ -c +\mu k + c\int dz\ d\tilde{z} P(z,\tilde{z})
    \left[ 1+ e^{\mu z +\tilde{z}}\right]^{-k} \right\} 
\end{equation}
For $k=O(\mu^{-1})$ in the limit $\mu\to\infty$, we find back the old 
saddle-point equation for the dominant effective chemical potentials.
The saddle-point equations for the sub-dominant corrections
$\rho^{(l)}(\tilde{z})$ are however obtained for $k=O(\mu^0)$. The
corresponding limit $\mu\to\infty$ is not obvious due to the existence
of terms like $\mu k$. We have to use equation (\ref{eq:Ba3}). The
left-hand side of the last equation thus reads
\begin{equation}
  \label{eq:Bi}
   \int dz\ d\tilde{z} P(z,\tilde{z}) e^{\mu z k +\tilde{z}k} = 
   \sum_{l=-1}^{\infty} p_l e^{-\mu k l} \int d\tilde{z}
   \rho^{(l)}(\tilde{z})  e^{\tilde{z}k}
\end{equation}
The dominant contribution for large $\mu$ and positive $k$ is given by
the term having $l=-1$, and diverges exponentially as $e^{\mu
  k}$. Multiplying equation (\ref{eq:Bh}) with $e^{-\mu k}$ thus
yields a well-defined limit $\mu\to\infty$, we find
\begin{equation}
  \label{eq:Bj}
  \rho^{(-1)}_{FT}(k) = \exp\left\{-cp_0+cp_0 \int d\tilde{z}
  \rho^{(0)}(\tilde{z}) \left( 1+e^{\tilde{z}} \right)^{ik} \right\}\ .
\end{equation}
We now proceed by subtracting the dominant contributions 
$\sim e^{\mu k}$ on both sides of eq. (\ref{eq:Bh}), and find for 
$\mu\to\infty$
\begin{eqnarray}
  \label{eq:Bk}
  \rho^{(0)}_{FT}(k) &=& \exp\left\{-cp_0+cp_0 \int d\tilde{z}
  \rho^{(0)}(\tilde{z}) \left( 1+e^{\tilde{z}} \right)^{ik} \right\}
  \rho^{(-1)}_{FT}(-k) \nonumber\\
  &=& \rho^{(-1)}_{FT}(k)\ \rho^{(-1)}_{FT}(-k)
\end{eqnarray} 
where we have used (\ref{eq:Bj}) in the last line. Continuing
by iteration, we finally find
\begin{equation}
  \label{eq:Bl}
  \rho^{(l)}_{FT}(k) =  \rho^{(-1)}_{FT}(k)\ \rho^{(-1)}_{FT}(-k)^l\
  . 
\end{equation}
So it is very simple to solve all but one of these saddle point
equations. We can consequently express $s_{VC}(x_c(c))$ in terms of
$\rho^{(0)}(\tilde{z})$, as is done in eq. (\ref{eq:smin}) in section 
\ref{sec:sc}. $\rho^{(0)}(\tilde{z})$ itself is described by
(\ref{eq:rho0}) which follows directly from equations
(\ref{eq:Bj},\ref{eq:Bk}). The corresponding calculations are lengthy
but straight-forward, so we do not present it here. The only trick
which has to be used is the following: Using (\ref{eq:Bl}) we may write
\begin{eqnarray}
  \label{eq:Bm}
  P_{FT}(k,\tilde{k})&=&\sum_{l=-1}^\infty\frac{W(c)^{l+2}}{(l+1)!\ c} 
  e^{ikl} \rho^{(l)}_{FT}(\tilde{k}) \nonumber\\
  &=& \frac{W(c)}{c} e^{-ik}\rho^{(-1)}_{FT}(\tilde{k})
  \sum_{l=-1}^\infty \frac{1}{(l+1)!} 
  \left[W(c)e^{ik}\rho^{(-1)}_{FT}(-\tilde{k}) \right]^{l+1}
  \nonumber\\
  &=& \frac{W(c)}{c} e^{-ik}\rho^{(-1)}_{FT}(\tilde{k})\ \exp\left\{
  W(c)e^{ik}\rho^{(-1)}_{FT}(-\tilde{k}) \right\}\ .
\end{eqnarray}
This expression helps to simplify $\ln P_{FT}(k,\tilde{k})$ in
equation (\ref{eq:Bf}).

\section{Evaluation of the RSB saddle-point equation}\label{app:rsb}

This last appendix shows how the $\mu\to\infty$ limit can be taken in
the one-step RSB saddle-point equation. We start with the order
paramter as given in (\ref{eq:1rsb})
\begin{equation}
  \label{eq:1rsbC}
  c[\nu] = \int {\cal{D}}\rho\ {\cal{P}}[\rho]\ \exp\left\{
  \int_0^1 dy\ \nu(y) \ln  \left[ 
  \int  dh\ \rho(h)\ \frac{e^{hmy}}{(1+e^h)^m} \right] \right\}
\end{equation}
and plug in ansatz (\ref{eq:rhoofh}),
\begin{equation}
  \label{eq:rhoofhC}
  \rho(h) = \omega_\mu \sum_{l=l_-}^{l_+} \rho_l e^{\mu m |l|/2} 
  \delta(h+\mu l)\ .
\end{equation}
We assume in particular that $\rho_{l_\pm}\neq 0$ for uniqueness
of the definition of $l_-$ and $l_+$.
Setting $\rho_l=0$ for all $l<l_-$ and all $l>l_+$, we find for the 
exponent in (\ref{eq:1rsbC}):
\begin{eqnarray}
  \label{eq:expC}
  \{...\}&=&\int_0^1 dy\ \nu(y) \ln  \left[ \int  dh\ \rho(h)\ 
  \frac{e^{hmy}}{(1+e^h)^m} \right]\nonumber\\
  &\simeq& \int_0^1 dy\ \nu(y) \ln  \left[\omega_\mu \rho_{-1} 
    e^{\mu m (y-1/2)} +\omega_\mu\frac{\rho_0}{2^m} + \omega_\mu
    \sum_{l>0} \rho_l e^{-\mu m l (y-1/2)}
  \right] 
\end{eqnarray}
where only the dominant contribution in $\mu$ is kept in every term
of $[\cdots]$. $\omega_\mu$ can be skipped in the last line, because
$\nu(y)$ has to have zero integral due to (\ref{eq:nuint}).
For large $\mu$ this is exponentially dominated by only one term which
depends on $y$: If $y<1/2$ the term with $l=l_+$ dominates, for
$y>1/2$ the $l_-$-term becomes exponentially larger than all others.
Introducing
\begin{eqnarray}
  \label{eq:nupmC}
  \nu_-&=&m\int_{1/2}^1 dy\ \nu(y)\ (\frac{1}{2}-y) \nonumber\\
  \nu_+&=&m\int_0^{1/2} dy\ \nu(y)\ (\frac{1}{2}-y) 
\end{eqnarray}
we conclude
\begin{equation}
  \label{eq:expoC}
  \lim_{\mu\to\infty}\mu^{-1} \{...\} = \nu_- l_- + \nu_+ l_+\ ,
\end{equation}
and
\begin{equation}
  \label{eq:cnuC}
  \lim_{\mu\to\infty} c[\nu/\mu] =\sum_{-1\leq l_-\leq l_+} 
  {\cal{P}}_{l_-,l_+} e^{\nu_- l_- + \nu_+ l_+}
\end{equation}
which is the left-hand side of the saddle-point equation. On the
right-hand side, an integral similar to (\ref{eq:expC}) has to be
determined. Following exactly the same scheme as above, we find the
expression given in equation (\ref{eq:rsbsaddle}).

\vspace{4cm}

\newcommand{\captionHeuristicBad}
{A small sample graph with minimum vertex cover of size 3. The
  vertices belonging to the minimum $V_{vc}$ are dark. For this graph
  the heuristic fails to find the true minimum cover, because is
  starts by covering the root vertex, which has the highest degree 3.}

\newcommand{\captionExampleDivide}
{Example how the divide-and-conquer algorithm operates. Above the
  graph is shown. The vertex $i$ with the highest degree is
  considered. In the case it is {\em covered} (left subtree), 
all incident edges can   be removed. In case it is {\em uncovered},
(right subtree)
all neighbors have to be {\em covered} and all edges incident to the
neighbors can be removed. In both case, the graph may split into
several components, which can be treated independently by recursive
calls of the algorithm.}

\newcommand{\captionXCC}
{Phase diagram: Fraction $x_c(c)$ of vertices in a minimal vertex
cover as function of the average connectivity $c$. 
For $x>x_c(c)$, almost all graphs have covers with $xN$ vertices, 
while they have almost surely no cover for $x<x_c(c)$.  
The solid line shows the replica-symmetric result. The circles
represent the results of the numerical simulations. Error bars are
much smaller than symbol sizes. The upper bound of Harant is given by
the dashed line, the bounds of Gazmuri by the dash-dotted lines.  The
vertical line is at $c=e$. Inset: All numerical values were
calculated from finite-size scaling fits of $x_c(N,c)$ using functions
$x_c(N)=x_c+aN^{-b}$.  We show the data for $c=2.0$ as an example.}

\newcommand{\captionBCC}
{The total backbone size $b_{uncov}(c)+b_{cov}(c)$ of minimal vertex 
covers as a function of $c$. The solid line shows the
replica-symmetric result, the dotted ones are the two results of
one-step RSB. Numerical data are represented by the error 
bars. They were obtained from finite-size scaling fits similar to the 
calculation for $x_c(c)$. The vertical line is at $c=e$ where
replica symmetry breaks down. }

\newcommand{\captionConn}
{Distribution of connectivities $d$ for $c=2.0$.  We show the total
connectivity distribution, given by a Poissonian of mean $c$, as well
as results describing the minimal vertex covers. The total
distribution is divided into three contributions arising from the
vertices which either are not in the backbone ($0<\langle x\rangle<1$)
or which are in the covered/uncovered backbone ($\langle
x\rangle=0/1$). Analytical predictions are represented by the lines
(which are guides to the eyes only, connecting the results for integer
arguments), while the numerical results for $N=17,35,70$ are displayed
using the symbols.}

\newcommand{\captionSCC}
{Entropy of minimal vertex as function of the average connectivity
$c$. The solid line results from the Gaussian approximation 
described in section \ref{sec:sc} while numerical data are given by the
symbols with error bars. Each numerical result was obtained by an
extrapolation $N\to\infty$ via fitting a function  
$s_c(N)=s_c+\alpha N^{-\beta}$ to the data for each $c$.
The vertical line is at $c=e$. }

\newcommand{\captionnonbb}
{
Examples of smallest non-backbone graphs. Note that all graphs can 
be divided into connected vertex pairs and some supplementary edges
connecting different pairs. A similar structure is found also for
the full non-backbone subgraph at connectivities $c<e$.
}

\newcommand{\captionFracLargest}
{Fraction $f_{\max}=C_{\max}/(1-b_c)N$
of the largest component of the non-backbone subgraph from
  numerical calculations as a function of graph sizes $N$ up to size
  $N=560$. For $b_c(c)$ the numerical values were taken.
  In a double logarithmic plot, for
  connectivities smaller than the predicted threshold $c\simeq 1.434$,
  the function $f_{\max}(N)$ has a negative curvature, indicating that
  $f_{\max}$ converges towards zero. Thus, for small connectivities,
  the non-backbone does not percolate. For larger connectivities,
  $f_{max}(N)$ has a positive curvature, and fits of the form
  $f(N)=f_{\infty}+bN^{c}$ result in stricly positive values, here
  $f_{\infty}=0.17(1)$ ($c=1.6$) resp. $f_{\infty}=0.37(3)$
  ($c=2.0$). Hence, the non-backbone percolates.
}

\newcommand{\captionnonbbmag}
{ Numerical histograms of local average occupation numbers $\langle
x\rangle_{\mu\to\infty}$ of non-backbone vertices for average
connectivities $c=2.0$ and $8.0$. The upper distribution is perfectly
symmetric as predicted by theory for all $c<e$. The lower one shows an
obvious bias towards higher occupation. The effect becomes stronger
with increasing connectivity.  Please note also the existence of
pronounced peaks in both distributions. These result from small
non-backbone components or dangling ends of the giant cluster, {\it
e.g.} the peaks at $m=1/3,2/3$ appear in chains of four vertices
connected by three edges as given by the second graph in the previous
figure. The weight of these peaks decreases with increasing size of
the giant non-backbone component.
}

\begin{figure}[htb]
\begin{center}
\myscalebox{\includegraphics{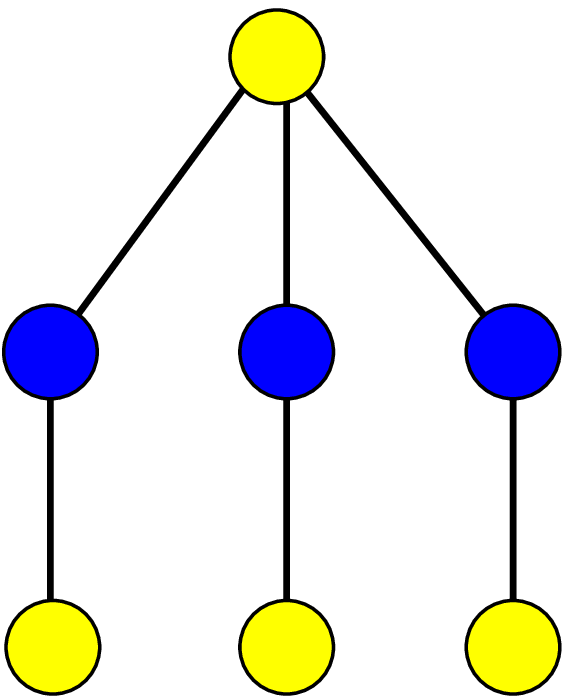}}
\end{center}
\caption{\captionHeuristicBad}
\label{figHeuristicBad}
\end{figure}

\begin{figure}[htb]
\begin{center}
\myscalebox{\includegraphics{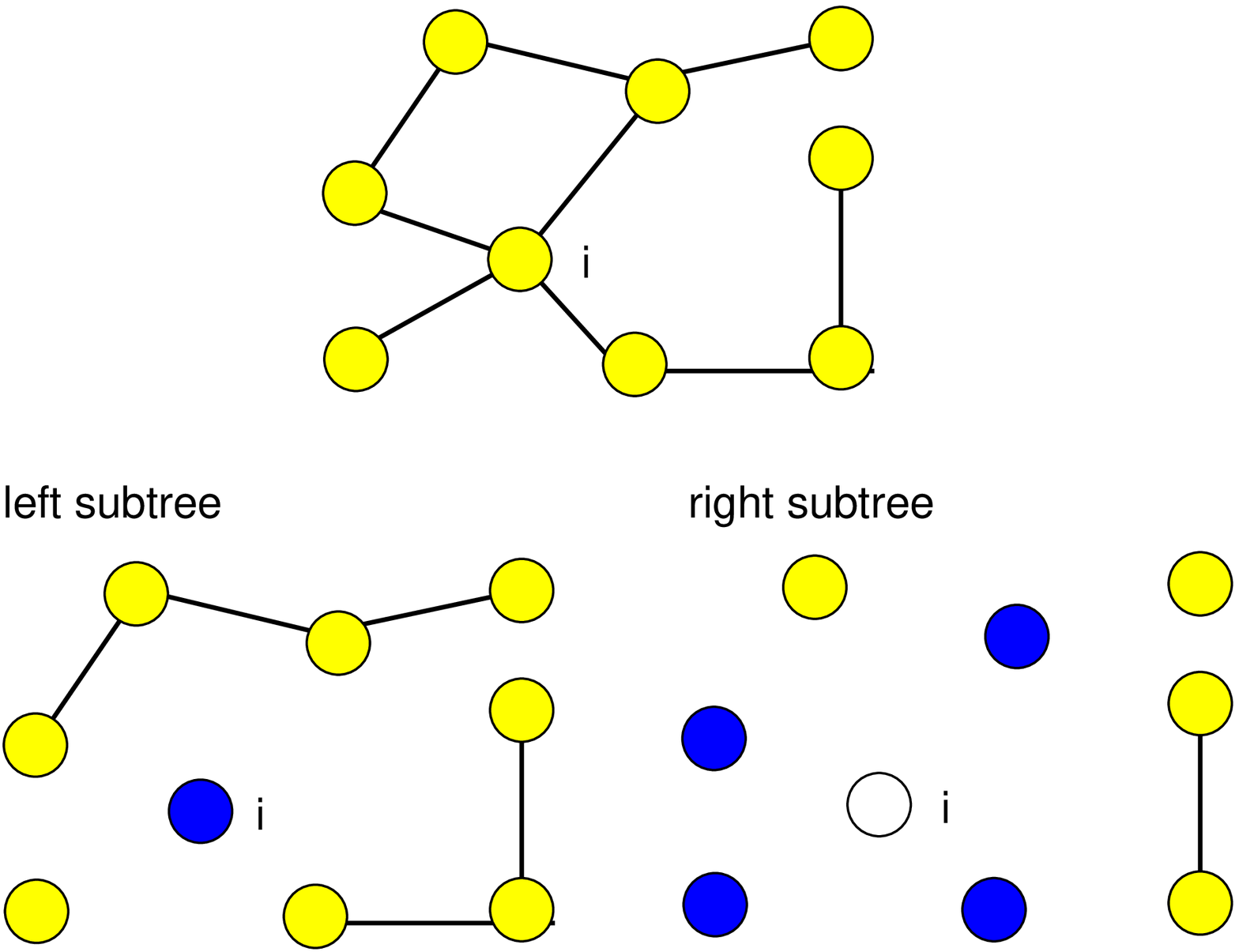}}
\end{center}
\caption{\captionExampleDivide}
\label{figExampleDivide}
\end{figure}

\begin{figure}[htb]
\begin{center}
\myscalebox{\includegraphics{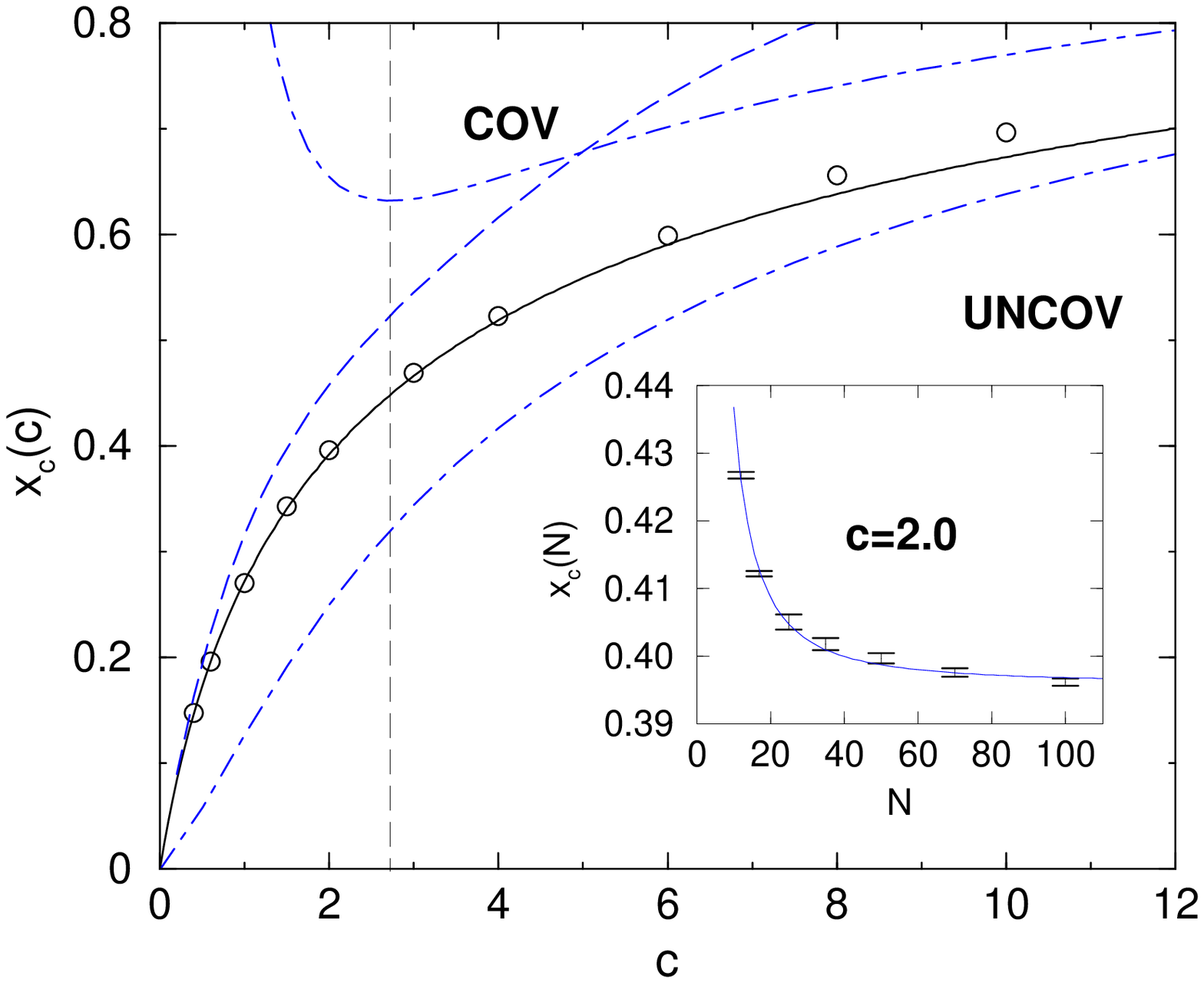}}
\end{center}
\caption{\captionXCC}
\label{figXCC}
\end{figure}

\begin{figure}[htb]
\begin{center}
\myscalebox{\includegraphics{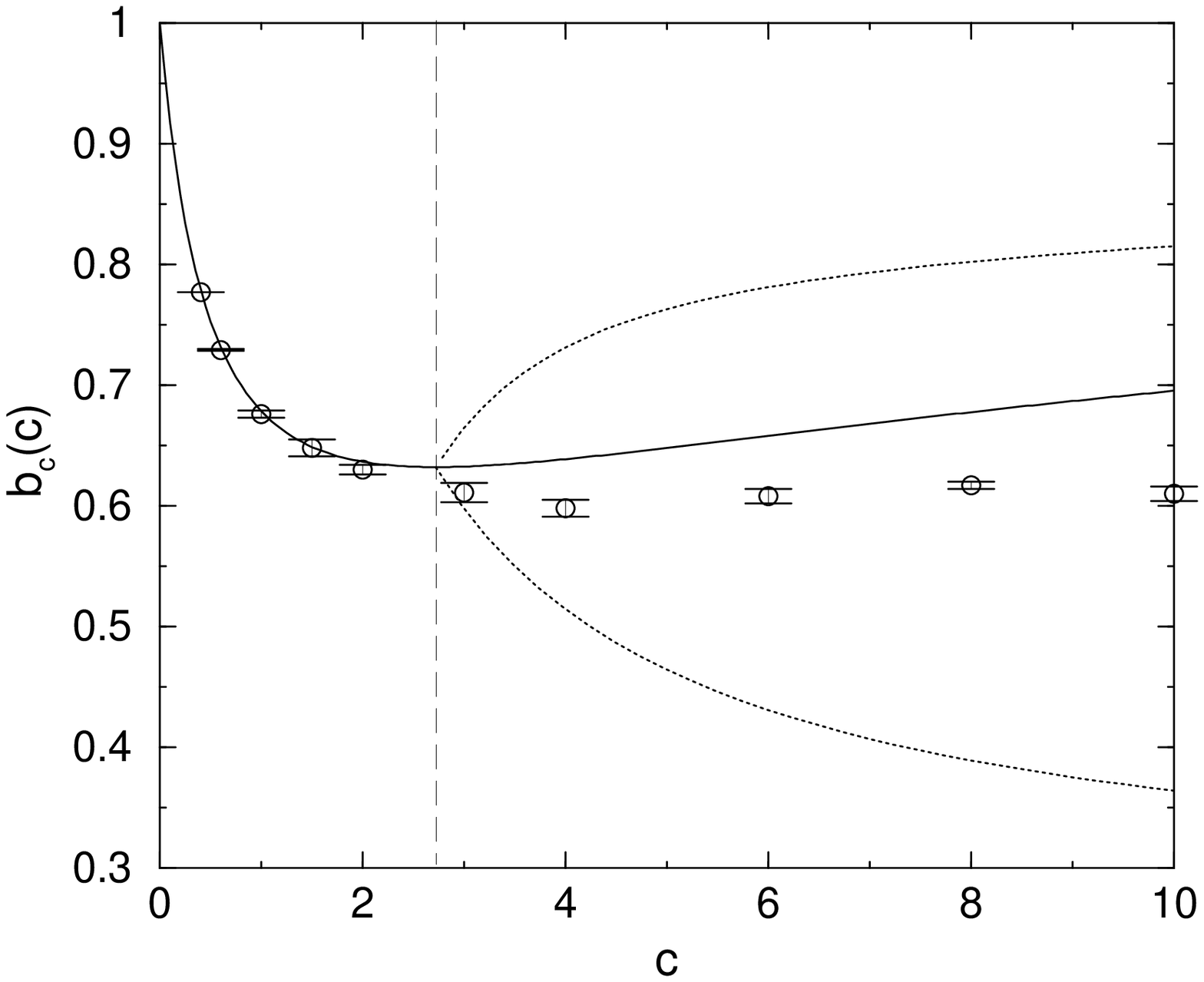}}
\end{center}
\caption{\captionBCC}
\label{figBCC}
\end{figure}

\begin{figure}[htb]
\begin{center}
\myscalebox{\includegraphics{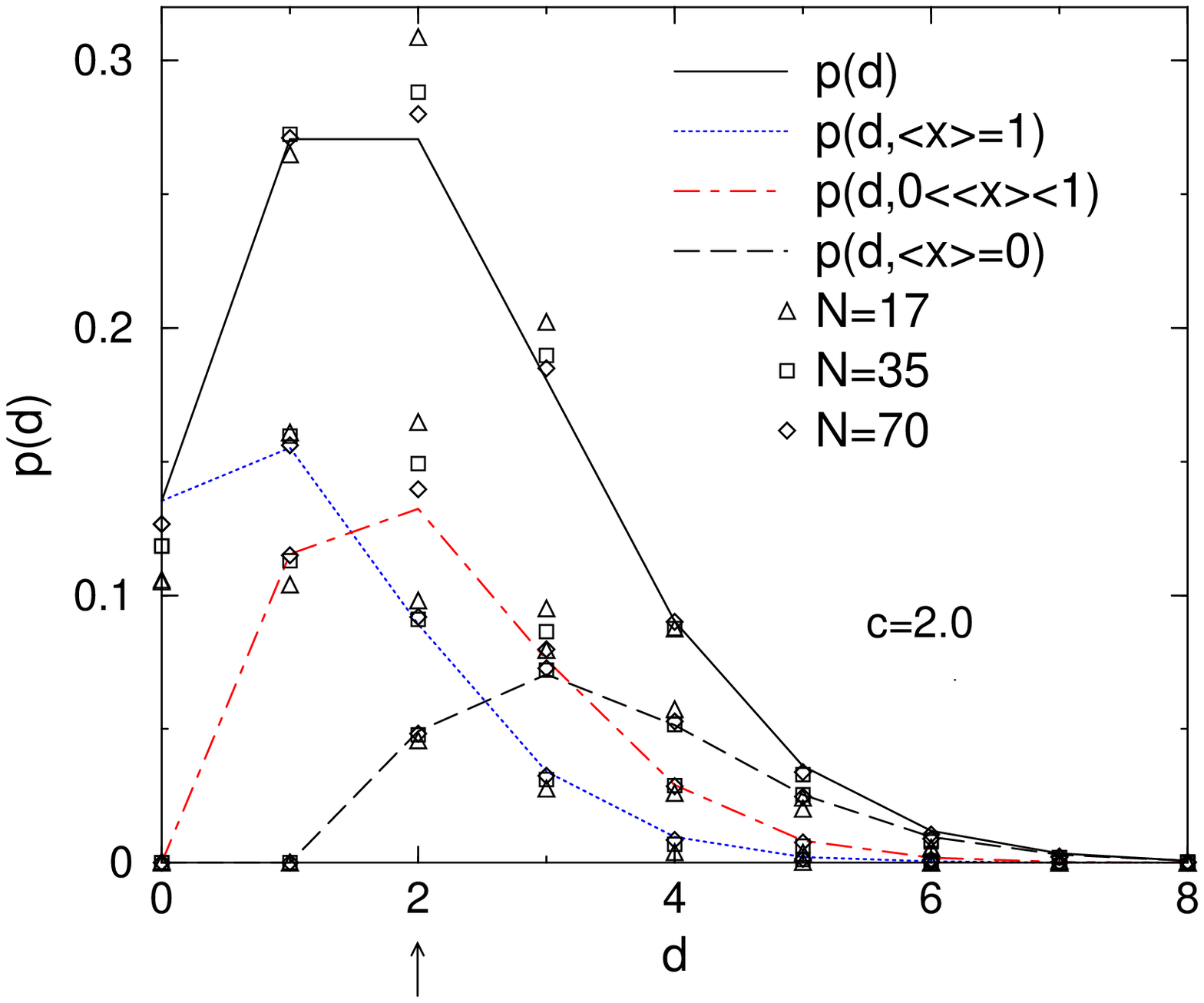}}
\end{center}
\caption{\captionConn}
\label{figConn}
\end{figure}

\begin{figure}[htb]
\begin{center}
\myscalebox{\includegraphics{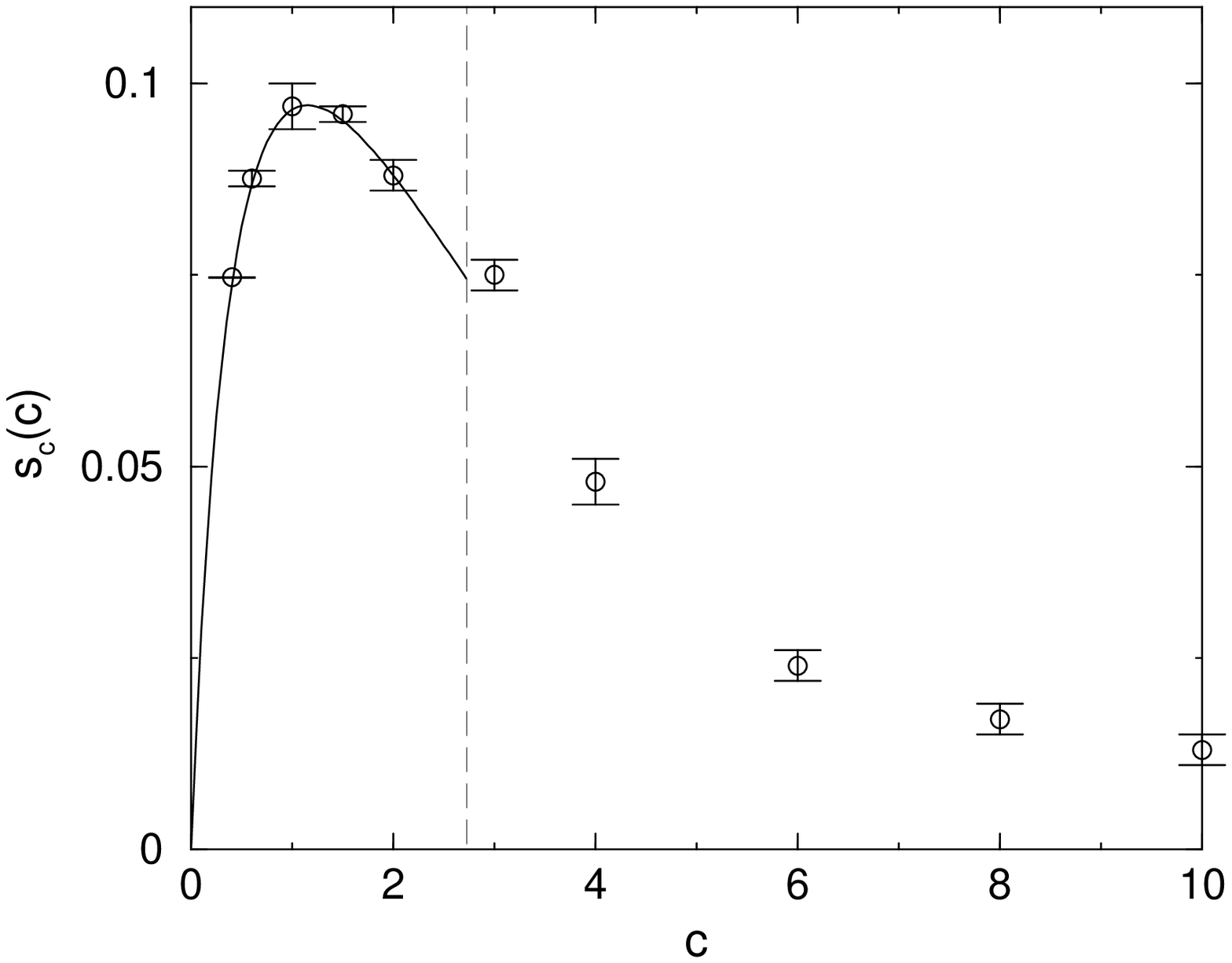}}
\end{center}
\caption{\captionSCC}
\label{figSCC}
\end{figure}

\begin{figure}[htb]
\begin{center}
\myscalebox{\includegraphics{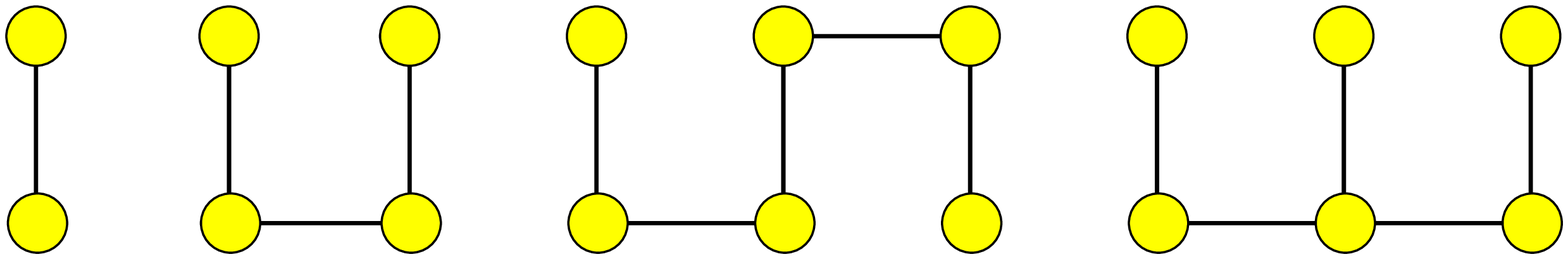}}
\end{center}
\caption{\captionnonbb}
\label{fig:nonbb}
\end{figure}

\begin{figure}[htb]
\begin{center}
\myscalebox{\includegraphics{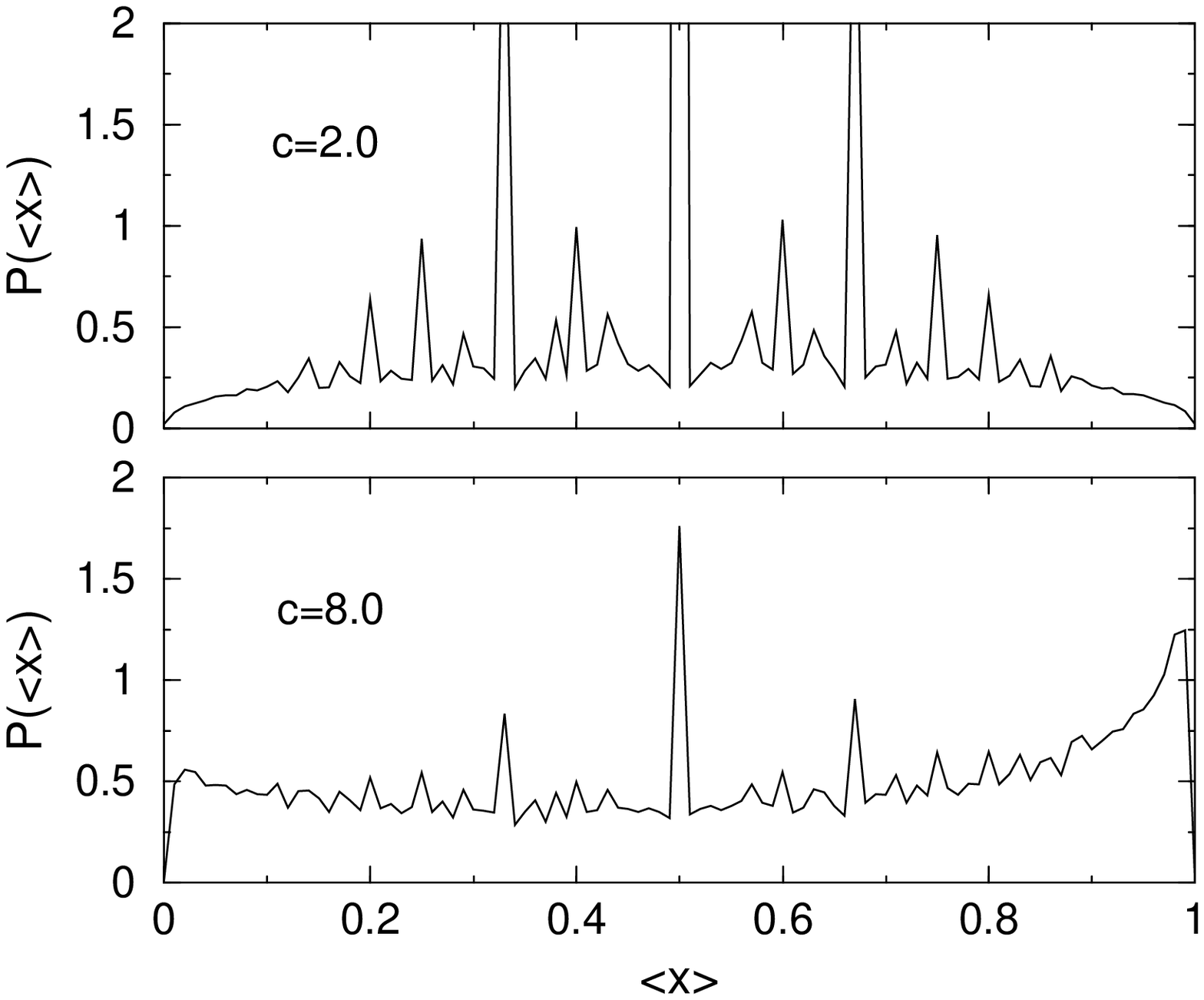}}
\end{center}
\caption{\captionnonbbmag}
\label{fig:magnonbb}
\end{figure}

\begin{figure}[htb]
\begin{center}
\myscalebox{\includegraphics{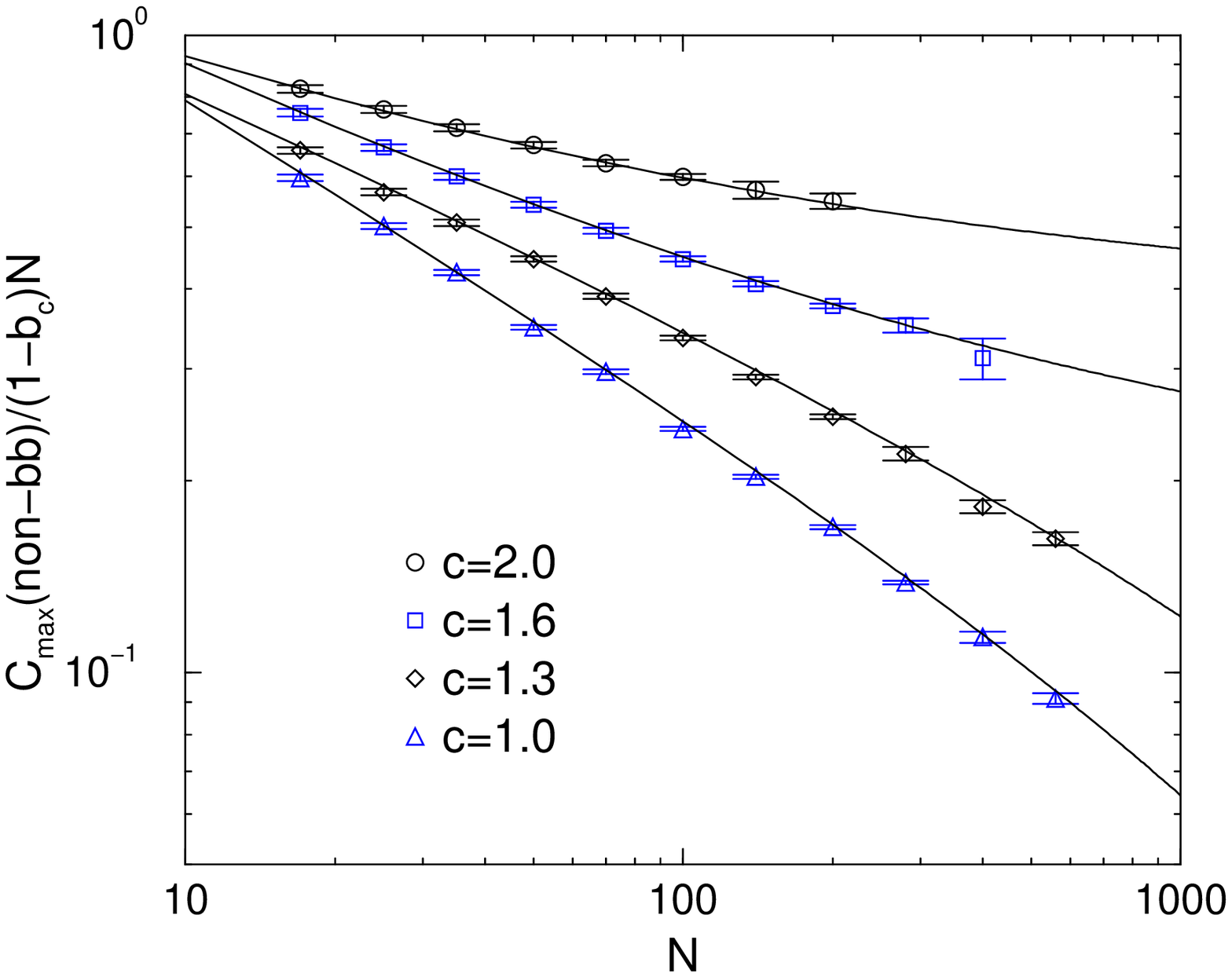}}
\end{center}
\caption{\captionFracLargest}
\label{figFracLargest}
\end{figure}


\begin{thebibliography}{99}
\bibitem{note} In principle, also a scaling $m\sim\mu^{-1}$ can be
  plugged in. This would be analogous to infinite-connectivity
  spin-glass models in the zero-temperature limit, and was also
  considered in the variational approach \cite{BiMoWe} to
  3-satisfiability. The inclusion into the presented formulation leads
  however to saddle-point equations which could not be solved by the
  authors.
\end{thebibliography}
\end{document}